\documentclass[10pt,aps,pra,longbibliography,showpacs,superscriptaddress,preprint,citeautoscript]{revtex4-1}

\usepackage{amssymb,amsmath,amsfonts,bm} 
\usepackage{graphicx}				\usepackage{color} 
\usepackage{dcolumn}   
\usepackage{bm}        
\usepackage{url}

\usepackage{hyperref}
\usepackage{subfigure}

\def\eeq{\relax}
\def\beq#1#2\eeq{\begin{equation}\label{#1}#2\end{equation}}
\def\bal#1#2\eal{\begin{align}\label{#1}#2\end{align}}
\def\bse#1#2\ese{\begin{subequations}\label{#1}#2\end{subequations}}
\def\ba{\begin{aligned}}   \def\ea{\end{aligned}}

\def\XXint#1#2#3{{\setbox0=\hbox{$#1{#2#3}{\int}$}
\vcenter{\hbox{$#2#3$}}\kern-.5\wd0}}

\newcommand{\ii}{\ensuremath{\mathrm{i}}}

\DeclareMathOperator{\real}{Re}
\DeclareMathOperator{\imag}{Im}

\def\dd{\operatorname{d}} 

\def\rev#1{\textcolor{blue}{#1}}	   
\def\an#1{\textcolor{magenta}{#1}}

\newcommand{\circled}[1]{\raisebox{.5pt}{\textcircled{\raisebox{-.9pt} {#1}}}}

\begin{document}

\title{The Inverse Grating Problem: Efficient Design of Anomalous Flexural Wave Reflectors and Refractors
} 
\author{Pawel Packo}
\affiliation{Department of Robotics and Mechatronics, AGH - University of Science and Technology,
Al. A. Mickiewicza 30, 30-059 Krakow, Poland}
\author{Andrew N. Norris}
\affiliation{Mechanical and Aerospace Engineering, Rutgers University, Piscataway, NJ 08854-8058 (USA)}
\author{Daniel Torrent}
\email{dtorrent@uji.es}
\affiliation{GROC, UJI, Institut de Noves Tecnologies de la Imatge (INIT), Universitat Jaume I, 12071, Castell\'o, (Spain)}

\date{\today}

 \begin{abstract}
We present an extensive formulation of the inverse grating problem for flexural waves, in which the energy of each diffracted mode is selected and  the grating configuration is then obtained by solving  a linear system of equations. The  grating is designed as a lineal periodic repetition of a unit cell  comprising  a cluster of resonators attached at points   whose physical properties are directly derived by inversion of a given matrix. Although both active and passive attachments can be required in the most general case, it is possible to find configurations with only passive, i.e.\ damped,  solutions.  This inverse design  approach presents  an alternative to the design of metasurfaces for flexural waves overcoming the limitations of gradient phase metasurfaces, which require  a continuous variation of the surface's impedance. When the grating is designed in such a way that all the energy is channeled to a single diffracted mode, it behaves as an anomalous refractor or reflector. The negative refractor is  analyzed in depth, and it is shown that with only three scatterers per unit cell is it possible to build such a device with  unitary efficiency.


 \end{abstract}

\maketitle

\section{Introduction}\label{sec1}

The fundamental property of gratings to redirect wave energy into multiple diffracted modes, transmitted and reflected, follows from simple considerations of interference effects.  This can be seen using ray theory for the incident and diffracted directions combined with the unit spacing on the grating: diffraction modes correspond to multiples of $2\pi$ in the phase difference of the incident and diffracted modes.  However, the related multiple  scattering problem of calculating the distribution of diffracted wave energy among the modes is far more difficult, and the inverse problem of selecting a desired energy distribution among these orders has been scarcely considered so far. Recently, some approaches based on complex acoustic and electromagnetic scatterers\cite{ra2017metagratings,epstein2017unveiling,wong2018perfect,quan2018maximum,rabinovich2018analytical,epstein2018perfect} have been proposed for  the design of gratings in which the energy is channeled towards a given direction.  This provides an interesting alternative method to overcome the limitations of gradient metasurfaces\cite{yu2011light}, in which a continuous variation of the phase at the interface is required to accomplish the directional channeling.  However, despite the recent interest in metagratings a systematic method for the design of gratings with specific energy distribution between modes has so far not been presented.

Recently, Torrent\cite{torrent2018acoustic}  considered a general acoustic reflective grating and   derived a linear relation between the grating parameters and the amplitudes of the diffracted orders. By selecting the diffracted amplitudes it is easy to obtain the grating parameters and  therefore to solve the inverse problem. In this specific case drilled holes in an acoustically rigid surface were selected as the basic grating elements. The purpose of this work is to demonstrate that a similar inverse design approach may be applied to flexural waves in thin plates. Here the grating comprises a one dimensional periodic repetition of a cluster of point attachments and the objective is to choose the number of these per unit cell  and their mechanical parameters (effective impedance) in order to control the diffracted wave amplitudes. 

The scattering of flexural waves by point attachments and compact inhomogeneities and its applications have been widely studied in the literature. Plane wave scattering from an array of finite points, an infinite line of equally spaced points, and from two parallel arrays is considered in \cite{Evans2007}.  Extensions to doubly infinite square and hexagonal arrays can be found  in \cite{Xiao2012} and \cite{Torrent2013}, respectively.  The hexagonal array introduces the possibility of Dirac cones in the dispersion surface, with implications for one-way edge waves \cite{Torrent2013,Pal2017}. A  method for dealing with wave scattering from a stack of gratings, comprising parallel gratings with pinned circles in the unit cell, is given by \cite{Movchan2009} and used to examine trapped modes in stacks of two \cite{Movchan2009} and three \cite{Haslinger2011} gratings.  The  scattering solution for a single grating is expressed in terms of reflection and transmission matrices, and recurrence relations are obtained for these matrices in the presence of a stack. Semi-infinite grating have recently been studied  \cite{Haslinger2017}.  The addition of point scatterers to  plates can produce flexural metamaterials with double-negative density and stiffness effective properties \cite{Gusev2014,Torrent2014}.  
Scattering from a 2D array of  perforations in a  thin plate designed to give high directivity for the transmitted wave is considered in \cite{Farhat2010a}.  Scattering of a Gaussian beam from a finite array of pinned points is examined in \cite{Smith2012}.   Time domain solutions of flexural wave scattering from platonic clusters  is considered in \cite{Meylan2011}.   Infinite arrays of  wave scatterers involve lattice sums for flexural waves, which, as we will see,  is relevant to the present work.  Lattice sums have other implications, for instance,  in the context of an infinite  square array of holes where the sums represent the consistency conditions between the local expansions at an arbitrary perforation and for the hole in the central unit cell \cite{Movchan2007}, also known as Rayleigh identities. 

\begin{figure}[h!]  
	\centering
	\includegraphics[width=10cm]{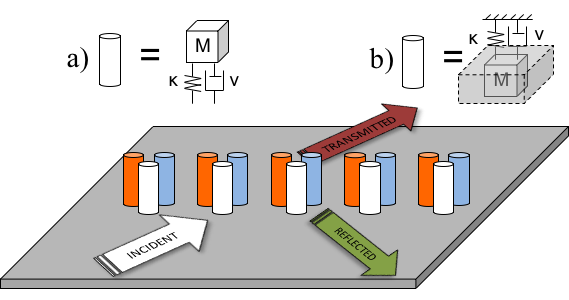}
	\caption{\rev{Schematic of the flexural wave grating problem: a line of clusters composed of multiple scatterers with possibly different mass, stiffness and damping properties within the cluster (but the same for each cluster). Each scatterer can be modeled as an attached  mass-spring-dashpot system with properly tuned constants $M$, $\kappa$ and $\nu$.  The schematic shows impedance models (a) and (b) of equation \eqref{-1}.} }
	\label{schematics2}
\end{figure}

In this work we \rev{consider an infinite array of point scatterers with the unit cell comprising a cluster of $N$ point scatterers characterized by scalar impedances.  A schematic of the grating and the  incident and scattered waves  is shown in Figure  \ref{schematics2}.  We focus on arrays of  periodically placed clusters with the intent of using the cluster properties to control forward and backward scattering.  Our approach is to first generalize the  forward scattering  methods of \cite{Evans2007} and \cite{Torrent2013}, and the derived expressions are then used to set up and solve the inverse grating problem.
Most importantly,   we note that \cite{Torrent2013} first presented a formalism for dealing with periodically arranged clusters of scatterers. This approach is the basis for the present work. }

The paper is organised as follows; section \ref{sec2} formulates the diffraction problem of a flexural plane wave by a periodic arrangement of clusters of $N$ scatterers. Section \ref{secinverse} defines the inverse grating problem and shows its solution and section \ref{applications} applies the theory to the design of a negative refractor. Finally, section \ref{summ} summarizes the work. Some mathematical results are derived in the Appendix.
	
\section{Diffraction by a periodic arrangement of point scatterers}\label{sec2}

\subsection{Scattering by a single and a cluster of point impedances}
The   deflection $w({\bf r})$ on a two dimensional plate, ${\bf r} = x \hat{\bf x} + y\hat{\bf y}$, satisfies the Kirchhoff plate equation
\beq{-2}
D\big( \Delta ^2 w({\bf r}) - k^4 w({\bf r}) \big) = 0 
\eeq
where 
$k^4 = \rho h\omega^2 /D$,  $D$ is the bending stiffness, $h$ is the plate thickness, and $\rho$ the  density. Time harmonic dependence $e^{-\ii \omega t}$ is assumed.  Equation \eqref{-2} holds   everywhere on the infinite plate  except where there are point impedances attached \cite{Evans2007}.  

Consider first scattering from a single point attachment   located at ${\bf r}={\bf R}$, 
\beq{-21}
D\big( \Delta ^2 w({\bf r}) - k^4 w({\bf r}) \big) = \mu w({\bf R}) 
\delta({\bf r}-{\bf R}) .
\eeq
The attached oscillator impedance $\mu$ is modeled as single degree of freedom with  mass $M$, spring stiffness  $\kappa$  and damping coefficient $\nu$.  Two possible models are 
\beq{-1}
\mu  = 
\begin{cases} 
\big( \frac 1{M\omega^2} - \frac 1{\kappa - \ii \omega \nu} \big)^{-1},
 & (a) ,
\\
M\omega^2 - \kappa + \ii \omega \nu , & (b)  .
\end{cases}
\eeq
In model (a) the mass is attached to the plate by a spring and damper acting in parallel \cite{Torrent2013}.  
Model (b) assumes the mass is rigidly attached to the plate, and both are attached to a rigid foundation by the spring and damper in parallel \cite{Evans2007}.  An important limit is a pointwise pinned plate, $w({\bf R})=0$, which corresponds to $\mu \to \infty$. 
\rev{The point attachments considered here are based on devices proposed for passive control of flexural waves using   {\it tuned    vibration absorbers} (TVA)s \cite{Brennan1997,Brennan1999,El-Khatib2005}. 
A TVA,  modeled as a point translational impedance, can be used  to reduce vibration at a specific frequency or to control transmission and reflection of flexural waves in a beam \cite{Brennan1999,El-Khatib2005}.  The  alternative  term {\it vibration neutralizer} \cite{Brennan1997}  is sometimes used.  In the present context the point impedance, or TVA, is considered as a device for controlling the scattering of flexural waves in two-dimensional  rather than in a 1D setting.}

The total plate deflection is 
\beq{1}
w({\bf r}) = w_\text{in}({\bf r}) + B G({\bf r}-{\bf R})
\eeq
where $w_\text{in}({\bf r})$ is the incident field and, by definition of the point impedance, 
\beq{2}
B = \mu w({\bf R}) . 
\eeq
Also,   $G$ is the Green's function (see Appendix \ref{A}) 
\beq{3}
G({\bf r} ) = C \big( H^{(1)}_0 (k r) - H^{(1)}_0 (\ii k r) \big) 
\eeq
where  $C = G({\bf 0} ) =\ii / (8k^2 D)$.  Note that  
$H^{(1)}_0 (\ii k r) = -\frac{2\ii}{\pi} K_0(kr)$. 
Setting ${\bf r}={\bf R}$ in \eqref{1} and using \eqref{2} yields 
\beq{4}
B = \frac{ w_\text{in}({\bf R}) }{\mu^{-1} - G({\bf 0} ) }. 
\eeq

If there are $N$ point scatterers located at 
${\bf R}_\alpha = x_\alpha \hat{\bf x} + y_\alpha \hat{\bf y}$ 
with impedances $\mu_\alpha$, $\alpha = 1, 2 , \ldots, N$, then the total field satisfies 
\beq{-214}
D\big( \Delta ^2 w({\bf r}) - k^4 w({\bf r}) \big) = \sum_{\alpha =1}^N \mu_\alpha w({\bf R}_\alpha) 
\delta({\bf r}-{\bf R}_\alpha) .
\eeq
The solution 
is given by the incident  field plus the field scattered
by all the particles, 
\beq{5}
w({\bf r}) = w_\text{in}({\bf r}) + \sum_{\beta = 1}^N B_\beta G({\bf r}-{\bf R}_\beta), 
\ \   B_\beta = \mu_\beta w({\bf R}_\beta) .
\eeq
Setting ${\bf r}={\bf R}_\alpha$ in \eqref{5} gives a linear system of $N$ equations for the amplitudes
\beq{6}
\sum_{\beta = 1}^N  
\big( \mu_\alpha^{-1} \delta_{\alpha \beta} - G({\bf R}_\alpha - {\bf R}_\beta )\big) 
B_\beta = w_\text{in}({\bf R}_\alpha) . 
\eeq

\subsection{Scattering by an infinite set of  impedances clusters   }

The above set of equations provides the solution for the multiple scattering problem of a given incident field on a cluster of small particles, once their position and their physical nature is properly
described. We would like to know what happens now when this cluster is copied and distributed along  a line and when the incident field is a plane
wave of definite wavenumber ${\bf k}$: 
\beq{7}
w_\text{in}({\bf r}) =   e^{\ii {\bf k}\cdot {\bf r}} . 
\eeq
 This defines the grating scattering problem. 

Specifically,  the grating particle positions are  
\beq{8}
{\bf R}_{\beta  m}= {\bf R}_\beta + {\bf R}_m
\eeq
where $\beta = 1, 2, ..., N$ defines the cluster element while ${\bf R}_m = m {\bf a}$, $m\in \mathbb Z$,  covers the infinite periodic grating. The total field is then
\beq{9}
w({\bf r}) = w_\text{in}({\bf r}) + \sum_{\beta = 1}^N \sum_{{\bf R}_m }
\mu_\beta  w({\bf R}_{\beta  m})G({\bf r}-{\bf R}_{\beta  m}). 
\eeq
It is assumed that the cluster-to-cluster relation for the total field satisfies the same phase relation as the incident field, 
\beq{10}
 w({\bf R}_{\beta  m}) = w({\bf R}_\beta ) e^{\ii {\bf k}\cdot {\bf R}_m}. 
\eeq
This crucial identity implies that the total field can be represented in terms of  $N$ amplitudes, $
\{B_\beta,\, \beta = 1,2,\ldots , N\}$, 
\beq{11}
w({\bf r}) =  e^{\ii {\bf k}\cdot {\bf r}}  + \sum_{\beta = 1}^N B_\beta \sum_{{\bf R}_m }
e^{\ii {\bf k}\cdot {\bf R}_m}
 G({\bf r}-{\bf R}_\beta - {\bf R}_m). 
\eeq
The amplitudes can be found by the same method as for the single cluster.   
Thus, 
setting ${\bf r}={\bf R}_\alpha$ in \eqref{11}  gives a linear system of $N$ equations
\beq{12}
\sum_{\beta = 1}^N  
\big( \mu_\alpha^{-1} \delta_{\alpha \beta} - 
\chi_{\alpha \beta} \big) B_\beta  =   e^{\ii {\bf k}\cdot {\bf R}_\alpha}
\eeq
with 
\beq{13}
\chi_{\alpha \beta} =
\sum_{{\bf R}_m }
e^{\ii {\bf k}\cdot {\bf R}_m} G({\bf R}_\alpha - {\bf R}_\beta - {\bf R}_m) .  
\eeq

\subsection{Solution of the forward scattering grating problem}
The $N$-cluster repeats along a line,  
\beq{14} 
{\bf R}_m = m a \hat{\bf x}, \ m \in \mathbb{Z},
\eeq
and therefore we can  use the lattice sum identity (see Appendix)
\bse{151}
\bal{15}
 \sum_{{\bf R}_m }
e^{\ii {\bf k}\cdot {\bf R}_m}
 G({\bf r}- {\bf R}_m) &=   G_0  \sum_{n\in \mathbb Z } e^{\ii (k_x+g_n) x  }
\Big( \frac{e^{-\zeta_-|y|}}{\zeta_-} - \frac{e^{-\zeta_+|y|}}{\zeta_+}
\Big), 
\\
G_0 = \frac 1{4Dk^2a} , \ \ 
g_n &= \frac{2\pi}a n, \ \ \zeta_\pm =((k_x+g_n)^2 \pm k^2)^{1/2} , 
\eal
\ese
where $\imag \zeta_- \le 0$. 
Specifically, \eqref{151} implies that the total field \eqref{11} is 
\beq{18}
w({\bf r}) =       e^{\ii {\bf k}\cdot {\bf r}}  + G_0 \sum_{\beta = 1}^N B_\beta
\sum_{n\in \mathbb Z }   e^{\ii (k_x+g_n) (x-x_\beta)  }
\Big( \frac{e^{-\zeta_-|y-y_\beta|}}{\zeta_-} - \frac{e^{-\zeta_+|y-y_\beta|}}{\zeta_+}
\Big)
\eeq
where the $N$ coefficients $B_\beta$ follow from eq.\ \eqref{12} and \eqref{13} with (instead of the general form   \eqref{6})
\beq{19}
\chi_{\alpha \beta } = G_0 
\sum_{n\in \mathbb Z } e^{\ii (k_x+g_n) (x_\alpha-x_\beta)  }
\Big( \frac{e^{-\zeta_-|y_\alpha-y_\beta|}}{\zeta_-} - \frac{e^{-\zeta_+|y_\alpha-y_\beta|}}{\zeta_+}
\Big). 
\eeq
This provides a much more computationally efficient expression than the  slowly convergent \eqref{13}. \rev{Note that the semi-analytical form for $\chi_{\alpha \beta } $ is a consequence of the fact that the Green's function can be expressed as a Fourier integral.  This indicates that the same procedure used in the Appendix would apply to other wave systems for which the  Green's function does not have a closed form solution. }

The $\zeta_+$ terms in the total field \eqref{18} all decay exponentially away from the line, while the $\zeta_-$ terms also decay except for those for which $\zeta_-$ is imaginary.  The latter define the finite set of {\it propagating modes}, ${\mathbb P}$ with $N_\text{P}$ elements, defined as
\beq{20}
 {{\mathbb P}} = \{ n \in {\mathbb Z}:\,  |k_x+g_n|<k\} .
\eeq
These are the values for which $\zeta_-$ is purely (negative) imaginary and they correspond to the far-field diffraction orders of the grating, all others are strictly near-field.  Note that 
${{\mathbb P}} $  always includes the value $n=0$, so that  $N_\text{P} \ge 1$. 

Let $\theta_0 \in [0,\frac{\pi}2 ]$ be the angle of incidence relative to the grating direction, so that 
\beq{8=3}
{\bf k} = k_x \hat{\bf x} + k_y\hat{\bf y}
= k\cos \theta_0  \hat{\bf x} + k\sin \theta_0 \hat{\bf y} . 
\eeq
In particular, $k_x = k \cos \theta_0$ implies that 
the direction of the propagating mode  $n $ is  defined by the angle 
\beq{-34}
\theta_n = \cos^{-1}\big( \cos \theta_0 + \frac{2\pi}{ka}n\big), \ \ 
\theta_n \in (0,\pi), \ \ 
n\in {\mathbb P}.
\eeq
Hence,  ${\mathbb P}$ can be considered as the set of $n$ for which $\theta_n$ is  real valued. 
The far-field diffracted displacement  is 
\beq{1211} 
w({\bf r}) =     e^{\ii {\bf k}\cdot {\bf r}}  + \frac{\ii G_0}k \sum_{\beta = 1}^N B_\beta
\sum_{n\in {\mathbb P} }  \frac 1{\sin \theta_n } e^{\ii k [ (x-x_\beta) \cos \theta_n 
+ |y-y_\beta| \sin \theta_n] }, 
\ \  |y|\to \infty .  
\eeq
The individual diffracted modes are therefore
\beq{22}
w({\bf r})=  
\begin {cases} \sum_{n\in {\mathbb P} }  
t_n 
  e^{\ii {\bf k}_n^+\cdot {\bf x}}, & y\to \infty ,
	\\ \sum_{n\in {\mathbb P} }  
	r_n
   e^{\ii {\bf k}_n^-\cdot {\bf x}},& y\to -\infty  ,  
\end {cases}
\eeq
where 
 ${\bf k}_n^+$, ${\bf k}_n^-$, are the wavenumbers of the transmitted  
and reflected waves, respectively, 
\beq{8=5}
{\bf k}_n^\pm 
= k\cos \theta_n  \hat{\bf x} \pm  k\sin \theta_n \hat{\bf y} , \ \ n \in \mathbb{P} 
\eeq
and the $2N_\text{P}$ transmission and reflection coefficients follow from \eqref{1211} and \eqref{22} as 
 \beq{45}
 \left. \begin{matrix}
t_n - \delta_{n0} \\  r_n
\end{matrix} \right\} = 
\frac{\ii G_0}{k \sin  \theta_n }  \sum_{\beta = 1}^N B_\beta
\times
\begin{cases}
 e^{-\ii {\bf k}_n^+\cdot {\bf R}_\beta } , 
\\
 e^{-\ii {\bf k}_n^-\cdot {\bf R}_\beta } , 
\end{cases} \ \  n\in {\mathbb P}.
\eeq
Note that ${\bf k}_0^+ = {\bf k}$, the incident wavevector, and that conservation of energy requires
\beq{-33}
\sum_{n\in {\mathbb P}} \big( |r_n|^2 + |t_n|^2\big) \sin \theta_n 
\le \sin \theta_0
\eeq
with equality if the impedances $\mu_\alpha$ are all real valued (no damping). 

Finally, we note that if all the scatterers lie along a line parallel to the  $x-$axis, i.e.\
$y_\beta = b$ $\forall \beta$ for some $b$, then 
\beq{7-}
 t_n - \delta_{n0} = r_ne^{-\ii 2 kb\sin \theta_n}, \ \  n\in {\mathbb P}.
\eeq
The number of independent scattering coefficients is therefore greatly reduced. 
This redundancy has implications in the selection of scatterer positions for the inverse grating problem, considered next. 

\section{The inverse grating problem} \label{secinverse}
\begin{figure}[h!]
	\centering
	\includegraphics[width=10cm]{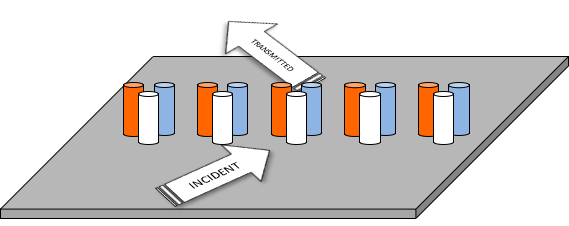}
	\caption{A grating which channels all of the wave energy into the $n=-1$ diffracted mode. } 
	\label{schematics}
\end{figure}

We are interested in controlling the reflection and transmission coefficients  through  (inverse) design of the grating.  \rev{For instance, Figure \ref{schematics} shows a grating that makes all but one of the scattered modes vanish, in this case all except the $n=-1$ mode. Specific designs for this type of grating are given below.}  The design and control is achieved using the combined degrees of freedom of the cluster spatial distribution, ${\bf R}_m$, the scatterers' positions, ${\bf R}_\alpha$, and their impedances, $\mu_\alpha$.  We consider the incidence direction $\theta_0$ and the nondimensional frequency $ka$ as given quantities.  The inverse problem as posed is still highly non-unique, since there could be multiple configurations that achieve the same objective.  We therefore    concentrate on specific geometrical configurations for the cluster distributions, such as a cluster of $N=3$ scatterers positioned at the vertices of a triangle or along a line.  This allows us to focus on the inverse problem of finding the impedances, and specifically on making them passive but with as little damping as possible so that all of the incident energy is channeled into the selected mode diffraction. 

\subsection{Inverting for impedances}
 
Equation \eqref{12}, written in  matrix form is 
\beq{12matrix}
 {{\bf E}} = {\bf M} {\bf B} 
\eeq
where the $N\times N$ matrix $\bf M$ follows from \eqref{12} and the $N-$vector $\bf E$ contains the 
incident wave amplitudes at the $N$ scatterer positions, 
\beq{100}
{\bf M} = {\pmb{\mu}}^{-1} - {\pmb{\chi}}, 
\ \  \
{\boldsymbol{\mu}} = 
\begin{pmatrix} \mu_1 &  &  & 0 \\
	  & \mu_2 &  &  \\
	  &  & \ddots & \\
	 0 &  & & \mu_N 
\end{pmatrix}, 
\ \  \
{\bf E} = \begin{pmatrix} 
e^{\ii {\bf k}\cdot {\bf R}_1} \\
e^{\ii {\bf k}\cdot {\bf R}_2 }\\ \vdots \\
e^{\ii {\bf k}\cdot {\bf R}_N}
\end{pmatrix} .
\eeq
The elements of the $N\times N$ matrix  $\pmb{\chi}$ are defined by the infinite sums \eqref{18}.  
Using the fact that $\pmb \mu$ is diagonal we can reconsider   \eqref{12matrix} as
an equation for $\pmb \mu$ in terms of the amplitudes $B_\alpha$, 
\beq{7=34}
\mu^{-1}_\alpha= \big( e^{\ii {\bf k}\cdot {\bf R}_\alpha }
+ {\bf e}_\alpha^T  {\pmb\chi} {\bf B} \big) 
/B_{\alpha}
\eeq
where the elements of the $N-$vector ${\bf e}_\alpha$ are zero except for the 
$\alpha^\text{th}$, which is  unity. 
In order to proceed we need to obtain the amplitudes ${\bf B}$. 


The goal is to control transmission coefficients, so we therefore 
 collect the transmission and reflection coefficients into a $2N_\text{P}-$vector denoted by ${\bf T} = [(t_n - \delta_{n0})\frac{\sin \theta_n}{\sin \theta_0}, r_n
\frac{\sin \theta_n}{\sin \theta_0}]^T$ with $n \in \mathbb{P}$. 
The vector length, $2N_\text{P}$, depends on the number of diffraction orders. Then, we may rewrite the equations for the transmission and reflection coefficients, \eqref{45}, as
\beq{26}
\begin{aligned}
	{\bf T} = {\bf S}{\bf B}
\end{aligned}
\eeq
with $ {\bf S}$, a $2N_\text{P} \times N$ matrix, collecting the exponential terms related to scatterer positions
\beq{27}
 {\bf S} = \frac {\ii G_0}{k \sin \theta_0} 
\begin{pmatrix} 
    e^{-\ii {\bf k}_0^+ \cdot {\bf R}_1 }
	& e^{-\ii {\bf k}_0^+ \cdot {\bf R}_2 } & \cdots & 	  
	  e^{-\ii {\bf k}_0^+ \cdot {\bf R}_N }            \\
	  e^{-\ii {\bf k}_0^- \cdot {\bf R}_1 }
	& e^{-\ii {\bf k}_0^- \cdot {\bf R}_2 } & \cdots & 
	  e^{-\ii {\bf k}_0^- \cdot {\bf R}_N }      \\
	  e^{-\ii {\bf k}_{-1}^+ \cdot {\bf R}_{1} } 
	& e^{-\ii {\bf k}_{-1}^+ \cdot {\bf R}_2 } & \cdots & 
	  e^{-\ii {\bf k}_{-1}^+ \cdot {\bf R}_N }            \\
	  e^{-\ii {\bf k}_{-1}^- \cdot {\bf R}_{1} } 
	& e^{-\ii {\bf k}_{-1}^- \cdot {\bf R}_2 } & \cdots & 
	  e^{-\ii {\bf k}_{-1}^- \cdot {\bf R}_N }      \\
	\vdots & \vdots & \vdots & \vdots \\
	  e^{-\ii {\bf k}_{n_P}^+ \cdot {\bf R}_1 }
	& e^{-\ii {\bf k}_{n_P}^+ \cdot {\bf R}_2 } & \cdots & 
	  e^{-\ii {\bf k}_{n_P}^+ \cdot {\bf R}_N }         \\
	  e^{-\ii {\bf k}_{n_P}^- \cdot {\bf R}_1 }
	& e^{-\ii {\bf k}_{n_P}^- \cdot {\bf R}_2 } & \cdots & 
	  e^{-\ii {\bf k}_{n_P}^- \cdot {\bf R}_N } 
\end{pmatrix}, 
\eeq
where $n_P$ indicates the $N_\text{P}^\text{th}$ diffracted mode.


We focus on the inverse grating problem of eliminating all but one of the $2N_\text{P}$  transmission and reflection coefficients.  Suppose we want all coefficients to vanish except, for instance, $t_m$ or $r_m$, then \eqref{26} provides 
$2N_\text{P} -1$ identities.   In order to have a solvable linear but not overdetermined system we require that the number of unknowns equals the number of knowns, implying a relation between the number of scatterers and the number of diffracted modes: 
\beq{9-4}
N = 2N_\text{P} -1. 
\eeq
The magnitude of the remaining coefficient must satisfy \eqref{-33}, implying 
\beq{2-6}
\begin{aligned}
	\hat{\bf T} = \hat{\bf S}{\bf B}
\end{aligned}
\eeq
where the $N-$vector $	\hat{\bf T}$ ($N= 2N_\text{P} -1$ vector) follows from ${\bf T}$ by removing the row for $t_m$ or $r_m$, and the square $N\times N$ matrix $\hat {\bf S}$ is obtained from the $2N_\text{P} \times 2N_\text{P} -1$ matrix $ {\bf S}$ 
by removing the row corresponding to the unconstrained coefficient ($t_m$ or $r_m$).   The $N$ scatterer amplitudes are therefore
\beq{28}
{\bf B}= {\hat{\bf S}}^{-1} \hat{\bf T} .
\eeq
It is important to note that we are assuming a non-singular ${\hat{\bf S}}$; the possibility and implications of ${\hat{\bf S}}$ being singular are discussed later. 
Substituting $\bf B$ into \eqref{7=34} yields the impedances in terms of the transmission/reflection vector $\hat{\bf T}$   as
\beq{7=35}
\mu^{-1}_\alpha=\frac{ e^{\ii {\bf k}\cdot {\bf R}_\alpha }
+ {\bf e}_\alpha^T  {\pmb\chi} \hat{\bf S}^{-1} \hat{\bf T} }
{ {\bf e}_\alpha^T  \hat{\bf S}^{-1} \hat{\bf T} }, \ \  \alpha = 1, 2, \ldots, N. 
\eeq
 
Equation \eqref{7=35} provides a simple inversion procedure at a given frequency for a given arrangement of scatterers  the number of which, $N$, is related to the number of diffraction orders, $N_\text{P}$, by equation \eqref{9-4}.  The latter implies that the number of scatterers is {\it odd}.   The solution \eqref{7=35} yields complex values for the impedances.  A realistic solution requires the further conditions that the impedances are passive, which is the case only if $\imag \mu_\alpha \ge 0$  $(\imag \mu_\alpha^{-1} \le 0)$ for all $\alpha = 1, 2 \ldots , N$.  

An explicit  solution follows for the case in which   all coefficients vanish except for the fundamental transmission $t_0$.  Then 
$\hat{\bf T} = {\bf 0}$ implying, from \eqref{7=35}, that $\mu_\alpha= 0$.  The solution is trivial: there is no grating.  For every other case, no matter which of the remaining $N = 2N_\text{P} -1$ coefficients is chosen as the one that is non-zero, the 
$N-$vector $\hat{\bf T} $ has the same form, {\it viz}.
\beq{4==}
\hat{\bf T} = (-1,\, 0, \, 0, \ldots , 0) = -{\bf e}_1.
\eeq
Equation \eqref{7=35} therefore simplifies to 
\beq{735}
\mu^{-1}_\alpha=\frac{  {\bf e}_\alpha^T  {\pmb\chi} \hat{\bf S}^{-1} {\bf e}_1 
-e^{\ii {\bf k}\cdot {\bf R}_\alpha }}
{ {\bf e}_\alpha^T  \hat{\bf S}^{-1}  {\bf e}_1 }, \ \  \alpha = 1, 2, \ldots, N. 
\eeq
In summary, if the impedances satisfy \eqref{735} then all but one of the transmission and 
reflection coefficients vanish. 

The matrix $\hat{\bf S}$ is invertible if and only if it is full rank, i.e.\ with $N$ linearly independent rows. If the scatterers are positioned along a line parallel to the $x-$axis, at the common coordinate $y_\beta = b$, then referring to \eqref{27}, 
$e^{-\ii {\bf k}_n^+ \cdot {\bf R}_\alpha }
= e^{-\ii {\bf k}_n^- \cdot {\bf R}_\alpha }
e^{-\ii 2kb\sin\theta_n }$. 
This implies that $\hat{\bf S}$  has at most $(N-1)/2$ linearly dependent rows, and therefore  the rank of the matrix falls precipitously from $N$ to 
$\frac 12 (N+1) = N_\text{P}$, see eq.\ \eqref{9-4}. Despite this singularity, it may happen that the expression \eqref{735} has a finite value by virtue of the fact this it contains $\hat{\bf S}^{-1} {\bf e}_1 $ in the numerator and in the denominator.  Also, $\hat{\bf S}^{-1} {\bf e}_1 $ itself can be finite even though $\hat{\bf S}$ is singular, as is the case in the example in Appendix \ref{B}.  Finally, the obvious exception to this discussion is the simplest, $N=1$, considered next.

\section{Examples and applications}\label{applications}

Following the theoretical developments for the inverse design of gratings, outlined in section \ref{secinverse}, we now present and discuss examples and applications. We   first focus on the simple case of $N = 1$, when only one diffracted mode exists, i.e. $n \in \mathbb{P} = \{ 0 \}$. Next, a more complex design for $N = 3$ (with $\mathbb{P}= \{ -1; 0 \}$) will be developed with particular focus on the inverse design of the cluster. This configuration   will be used to find scatterer configurations resulting in the negative refraction of waves at the grating. 

The negative refractor consists in a grating that diverts an incoming wave in such a way that if the angle the wave makes with the $x-$axis is $\theta_0$ that of the transmitted wave is $\pi -\theta_0$. This is indeed the ``refraction'' version of the retroreflector, in which the incident wave is retroreflected. From the diffraction point of view we assume that the selected incident angle $\theta_0$ allows for two diffracted modes $\mathbb{P} = \{ -1; 0 \}$. We  also want  the angle of the $n=-1$ mode to be  $\theta_{n}=\pi -\theta_0$, therefore, using $k = \frac{2\pi}{\lambda}$ equation \eqref{-34} gives 
\beq{diff}
-\cos\theta_0=\cos\theta_0-\frac{\lambda}{a}
\eeq
which sets up the ratio $\lambda/a=2\cos\theta_0$ (or $ka = \pi \sec \theta_0$). A configuration of $N=3$ scatterers will be used to demonstrate the negative refractor.

\subsection{The simple grating: $N=1$} 

By assumption, the fundamental is the only diffracted order and the only transmission/reflection  coefficients are related by $t_0 = r_0+1$, assuming with no loss in generality that it is positioned at ${\bf R}_1 = {\bf 0}$. Equation \eqref{735}, the condition for total reflection  $(t_0=0\, \Rightarrow \, r_0=-1)$,  reduces to a  scalar relation
\beq{9=3}
 \mu ^{-1} = \chi  - \frac {\ii G_0}{k\sin\theta_0} 
\eeq
where $\chi = \chi_{\alpha \alpha}$  follows from \eqref{19}. In particular \cite{Evans2007} since ${\mathbb P}=\{0 \}$, 
\beq{==5}
\chi-\frac {\ii G_0}{k\sin\theta_0} 
\equiv 
\chi_1 = \frac{-G_0}{k\sqrt{1 + \cos^2\theta_0}}
+ G_0 
\sum_{n\in \mathbb Z \backslash 0}
\Big( \frac 1{\sqrt{ (k_x+g_n)^2 - k^2}}- \frac 1{\sqrt{ (k_x+g_n)^2 + k^2}}
\Big)
\eeq
where $\chi_1$ is real. Total reflection can therefore be achieved with real impedance $\mu = \chi_1^{-1}$, a result previously obtained in \cite{Evans2007}. 

Since there is only one scattering coefficient in this case (because $t_0 = r_0+1$), it is of interest to see what other values of $t_0$ can be achieved.  Instead of using \eqref{4==} we retain $t_0\ne 0$ and set $\hat{T} = t_0-1$. 
Equation \eqref{7=35} then simplifies to 
\beq{9=5}
 \mu ^{-1} =\frac{\ii G_0 t_0}{k\sin\theta_0 (t_0-1)}  +\chi_1  .
\eeq
Equation \eqref{9=5} provides an explicit expression for the impedance for a given incidence direction $\theta_0$, lattice spacing $a$, wavenumber $k$ and transmission $t_0$.  The impedance is complex valued, indicating damping is necessary, except for the two limiting values  $t_0 = 0$, discussed above,  and  $t_0 = 1$ which is the trivial limit of $\mu = 0$, i.e.\  no grating.  

What other values of $t_0$ can be achieved with a {\it passive} impedance?  Recall that a passive impedance  maintains or  dissipates energy, as opposed to an active impedance which requires an external energy source. 
The impedance is passive iff $\imag \mu^{-1} \le 0$, e.g.\ see \eqref{-1}.   Equation 
\eqref{9=5} gives a passive $\mu$ iff $\real \frac {t_0}{t_0-1} \le 0$. Hence, 
\beq{0--2}
t_0 = |t_0| e^{\ii \phi}, \ \  |t_0| \le \cos\phi 
\ \Leftrightarrow \ \text{passive }\ \mu. 
\eeq
In addition to the limits  $t_0 = 1$ and $t_0=0$ discussed above, this provides the entire range of  transmission coefficients achievable with $N=1$.

\subsection{The next simplest grating: $N=3$} \label{3.2}
 
There are two diffracted modes, $\mathbb{P}= \{ -1; 0\}$, if the incidence angle $\theta_0$ is large enough, which is now assumed. Following the discussion in section \ref{secinverse} we will need three scatterers, $N = 3$, in order to control three out of four reflection and transmission coefficients. With the goal of designing a negative refractor we want the only propagating mode, among all transmitted through or reflected from the grating, be the transmitted $n = -1$ order. A grating that sends all of the incident energy into the transmitted $n=-1$ mode has matrix $ \hat{\bf S} $, 
\beq{27.1}
 \hat{\bf S} = \frac {\ii G_0}{k \sin \theta_0} 
\begin{pmatrix} 
    e^{-\ii {\bf k}_0^+ \cdot {\bf R}_1 }
	& e^{-\ii {\bf k}_0^+ \cdot {\bf R}_2 } &  
	  e^{-\ii {\bf k}_0^+ \cdot {\bf R}_3 }            \\
	  e^{-\ii {\bf k}_0^- \cdot {\bf R}_1 }
	& e^{-\ii {\bf k}_0^- \cdot {\bf R}_2 } &  
	  e^{-\ii {\bf k}_0^- \cdot {\bf R}_3 }      \\ 
	 e^{-\ii {\bf k}_{-1}^- \cdot {\bf R}_1 }
	& e^{-\ii {\bf k}_{-1}^- \cdot {\bf R}_2 } &  
	  e^{-\ii {\bf k}_{-1}^- \cdot {\bf R}_3 }     
\end{pmatrix}.
\eeq

\begin{figure}[h!]
	\centering
	\begin{subfigure}{}
		\includegraphics[width=0.48\linewidth]{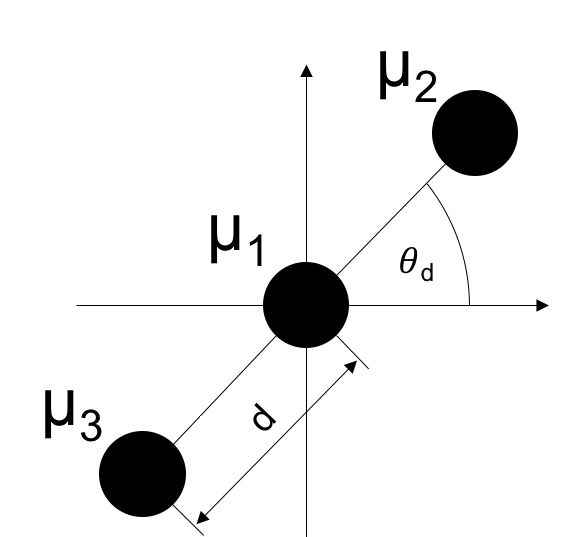}
	\end{subfigure}
	\begin{subfigure}{}
		\includegraphics[width=0.48\linewidth]{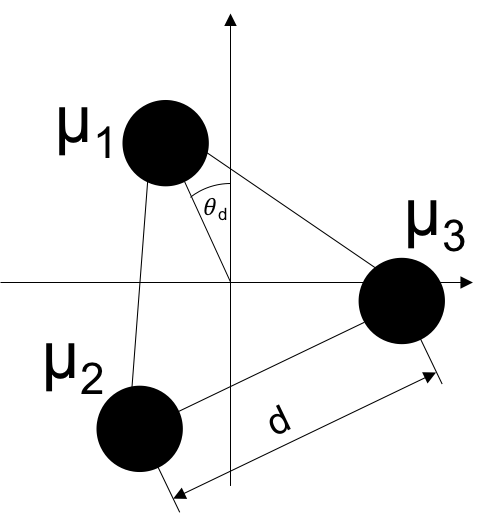}
	\end{subfigure}
	\caption{Two investigated cluster configurations: (a) linear and (b) triangular.}
	\label{N3clusters}
\end{figure}

We consider  two particular geometrical setups for $N = 3$ clusters, namely a linear and triangular cluster, as shown in Figure \ref{N3clusters}. In each case we parametrize the cluster by   the   spacing between the scatterers and the rotation angle of the cluster,   $d$ and   $\theta_d$, respectively. 
The positions of the scatterers in the cluster are then: ${\bf R}_1 = (0,0)$ and $\bm R_{2 \atop 3} = \pm d(\cos\theta_d,\sin\theta_d)$, $d > 0$ for the linear cluster; and ${\bf R}_{1} = \frac d{\sqrt{3}} \ \big( -\sin \theta_d  , \cos \theta_d \big)$ and ${\bf R}_{2 \atop 3} =  \frac d{\sqrt{3}} \ \big( \sin (\theta_d \mp \frac{\pi}6 ), - \cos (\theta_d \mp \frac{\pi}6 )\big)$ 
for the triangular cluster. 

The design process for a grating consists of finding scatterers' impedances and positions $(d,\theta_d)$. Among all possible solutions we are interested  in passive cluster configurations, i.e. $\imag \mu_\alpha > 0$ for all $\alpha$, that correspond to the largest possible transmission coefficient $|t_{-1}|$. The latter would imply that possibly large portion of energy of the incident wave is sent into the transmitted $n = -1$ mode, resulting in the negative refractor. From a practical perspective, a particularly interesting cluster setup would satisfy $\imag \mu_\alpha = 0$, resulting in spring-mass configurations of the scatterers only (no damping).

In the following examples we assume the incident wavevector $k = \pi / (a \ \cos \theta_0)$ at angle $\theta_0 = \pi/4$. In each case all but $t_{-1}$ reflection and transmission coefficients in $\hat{{\bf T}}$ are set to zero.  We also   assume, for simplicity,  $D = 1$ and $a = 1$.


\subsection{Numerical examples}

\subsubsection{Results for the linear cluster}

\begin{figure}[h!]
	\centering
	\begin{subfigure}{}
		\includegraphics[width=0.48\linewidth]{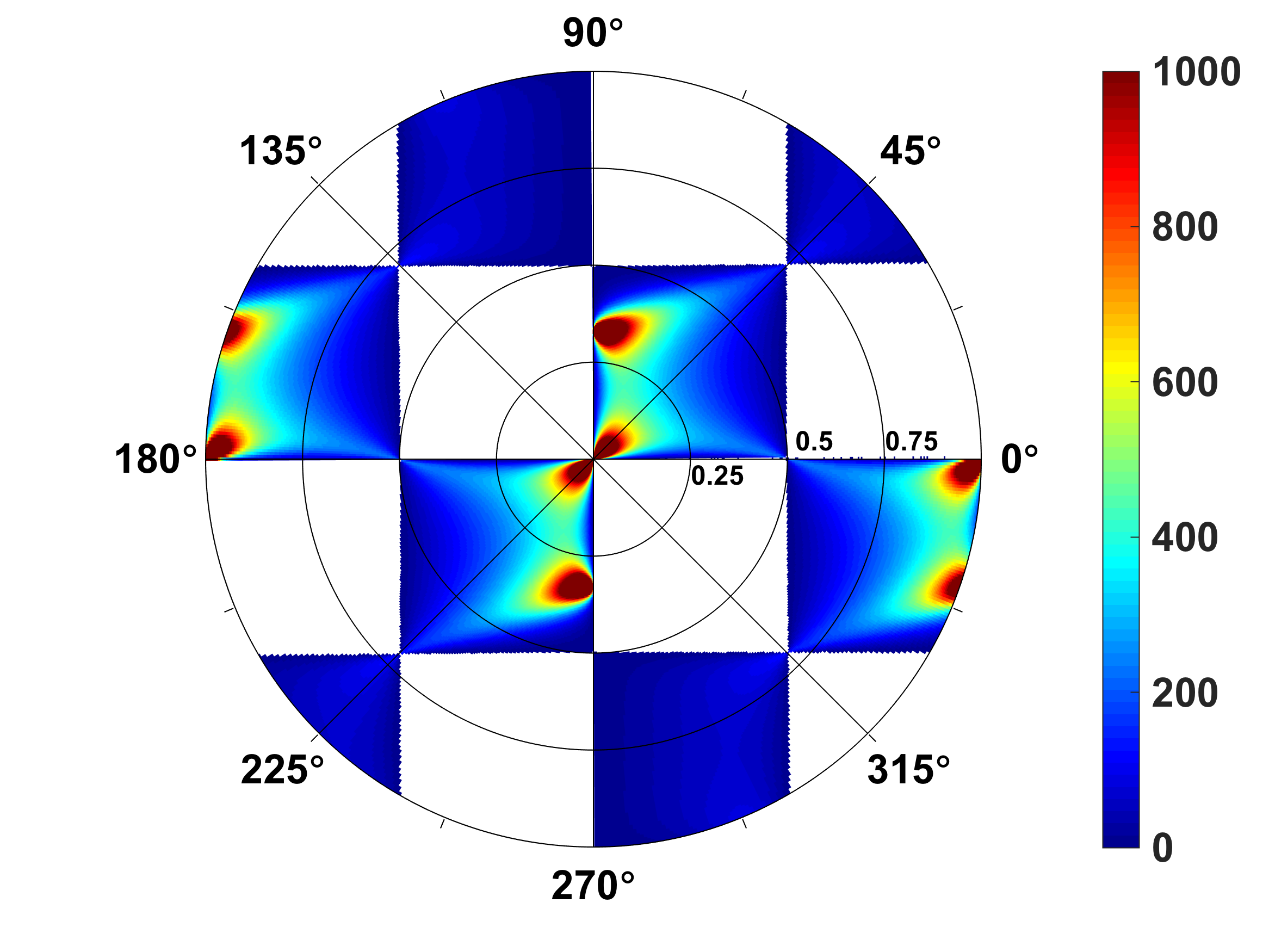}
	\end{subfigure}
	\begin{subfigure}{}
		\includegraphics[width=0.48\linewidth]{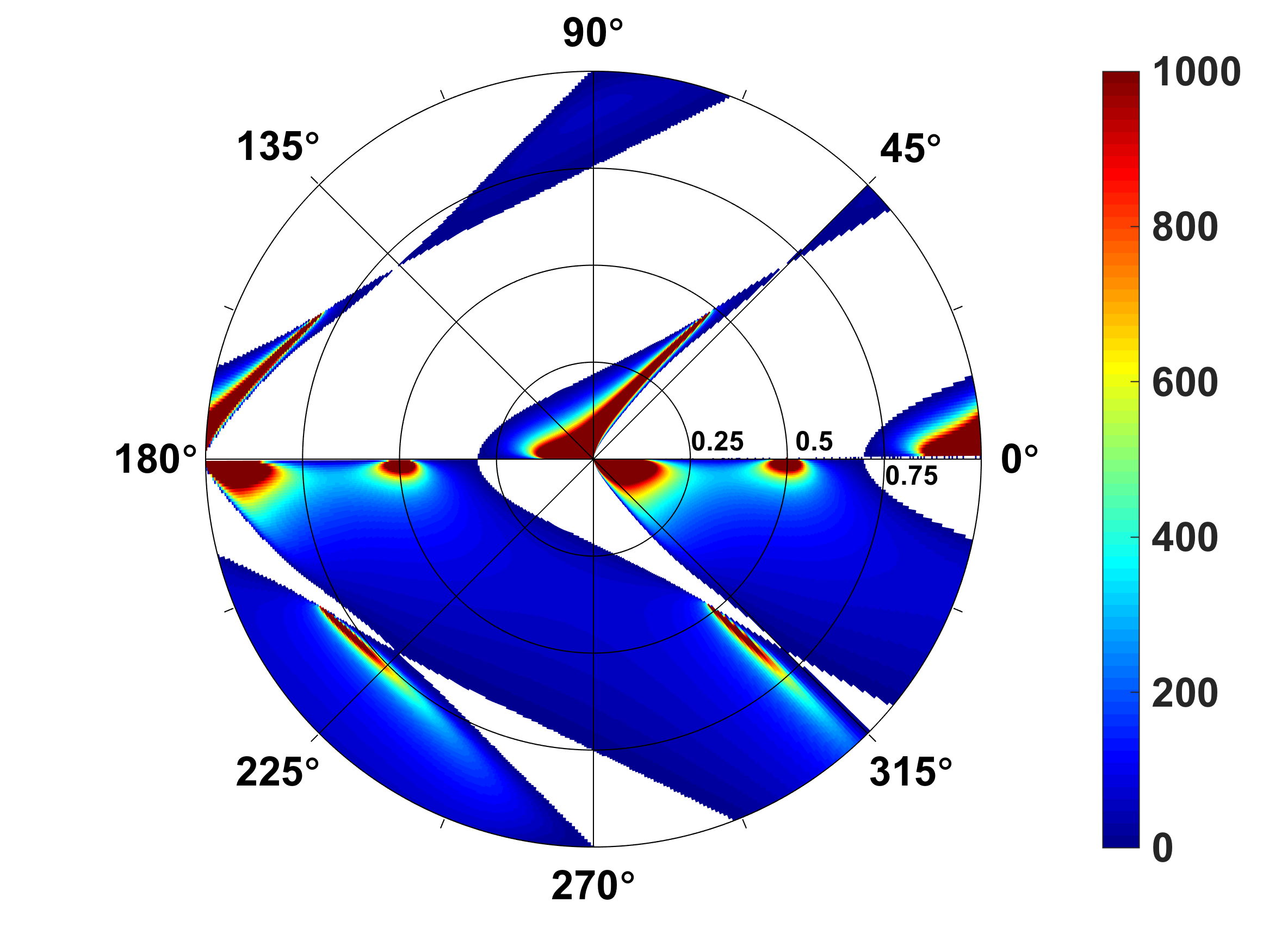}
	\end{subfigure}
	\caption{Imaginary parts of the complex-valued impedance $\mu_1$  (left) and $\mu_2$ (right), in units of $a^2/D$, as functions of $d$ and $\theta_d$ for a fixed incident wavevector angle $\theta_0 = \pi/4$ and $k = \pi / (a \ \cos \theta_0)$ (for $a = 1$), for the $N=3$ linear cluster negative refractor.  The plot only shows regions for which the impedances are  passive: $\imag \mu_1, \mu_2 \ge 0$.}
	\label{n3mu1mu2imag}
\end{figure}

\begin{figure}[h!]
	\centering
	\begin{subfigure}{}
		\includegraphics[width=0.48\linewidth]{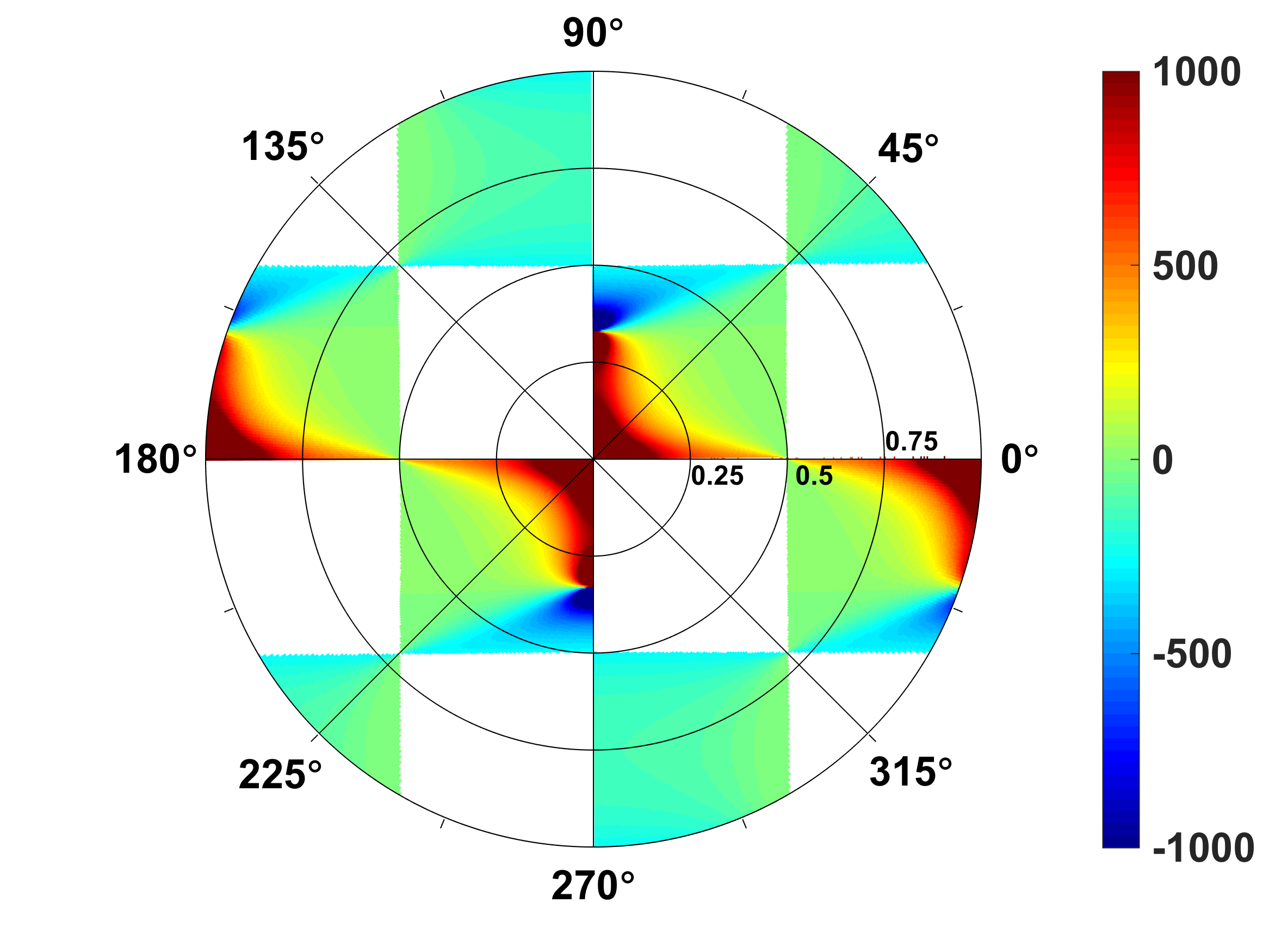}
	\end{subfigure}
	\begin{subfigure}{}
		\includegraphics[width=0.48\linewidth]{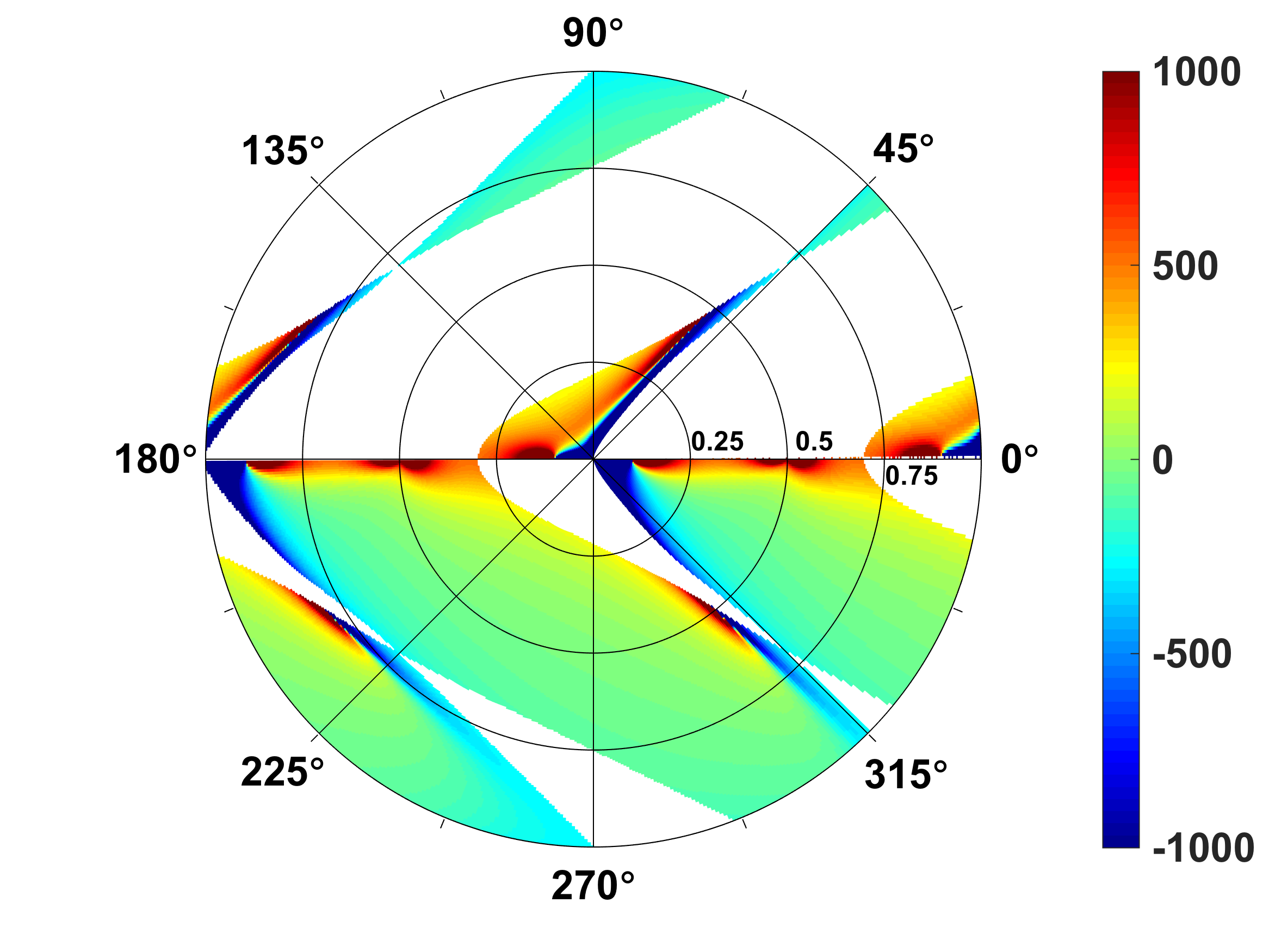}
	\end{subfigure}
	\caption{Real parts of the impedances $\mu_1$  (left) and $\mu_2$ (right), in units of $a^2/D$, as functions of $d$ and $\theta_d$ for   incident wavevector angle $\theta_0 = \pi/4$ and $k = \pi / (a \ \cos \theta_0)$ (for $a = 1$), for the $N=3$ linear cluster negative refractor.  The plot is restricted to passive impedances, see Fig.\ \ref{n3mu1mu2imag}.}
	\label{n3mu1mu2real}
\end{figure}

\begin{figure}[h!]
	\centering
	\includegraphics[width=0.8\linewidth]{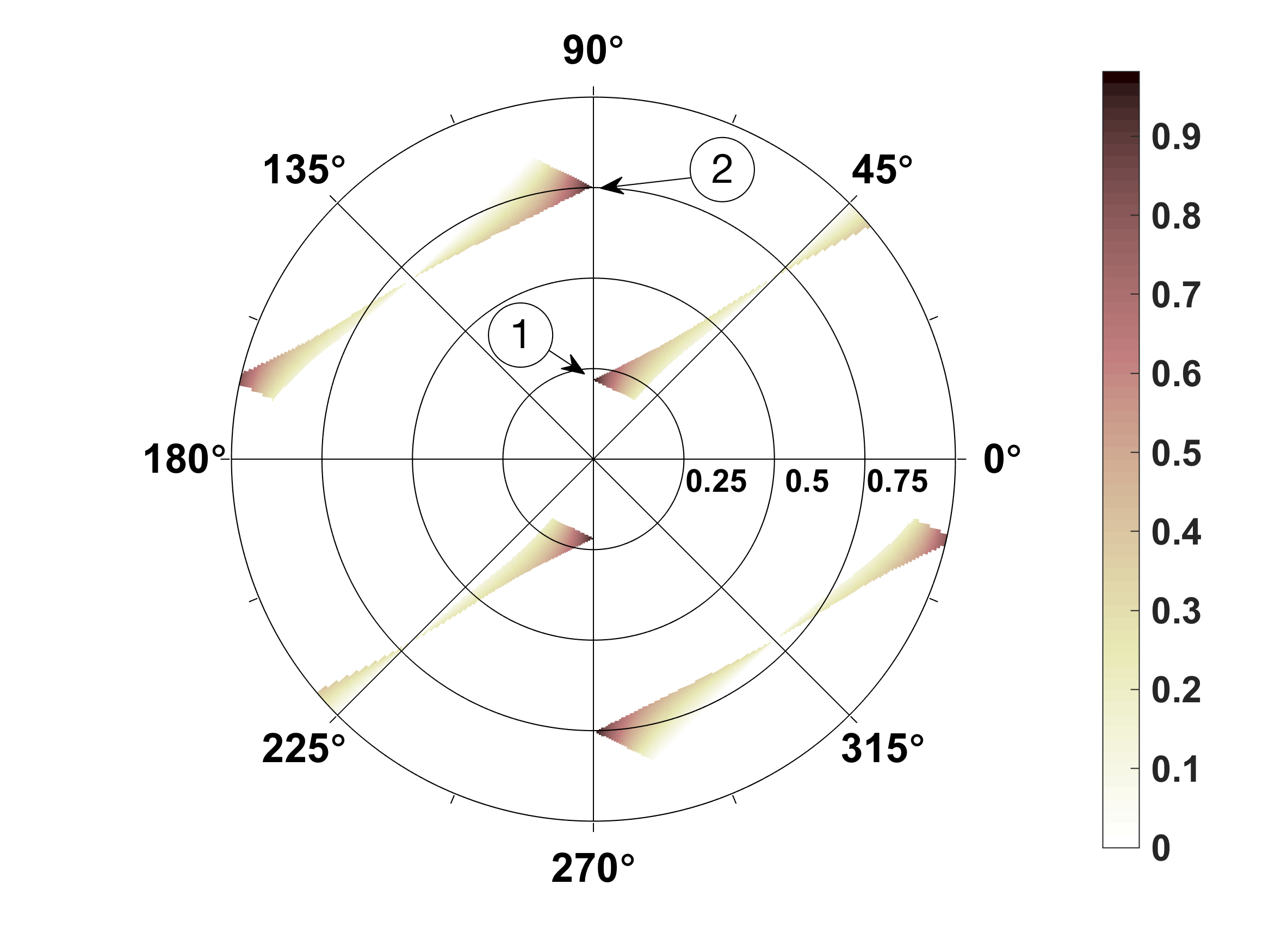}
	\caption{Magnitude of the transmission coefficient $t_{-1}$ for the designed linear $3-$cluster negative refractor as a function of $(d, \theta_d)$ plotted in  polar coordinates. The regions shown  corresponding to passive $\mu_\alpha$ for all $\alpha$, and labels on the  selected points are consistent with table \ref{tabclusters}. 
	 }
	\label{n3common}
\end{figure}

We begin by inverting $\hat{{\bf S}}$ from \eqref{27.1} and using \eqref{7=35} to solve for impedances. Figures \ref{n3mu1mu2imag} and \ref{n3mu1mu2real} show, respectively, the  imaginary and  real  parts of the complex-valued impedance $\mu_1$ and $\mu_2$ ($\mu_3$ is similar to $\mu_2$ due to the symmetry of the cluster) for $\theta_d \in (0, 2 \pi)$ and $d \in (0, a)$. As we are interested in passive solutions only, the plots in Figures  \ref{n3mu1mu2imag} and \ref{n3mu1mu2real} are limited to $(d,\theta_d)$ combinations resulting in $\imag \mu_\alpha > 0$ for respective scatterers independently.
A cluster with passive damping properties can only be constructed by selecting scatterers positions corresponding to impedances satisfying $\imag \mu_\alpha > 0$ for all $\alpha$. Those combinations of $(d, \theta_d)$, with the values of $|t_{-1}|$ are shown in Figure \ref{n3common}.

Cluster configurations corresponding to the highest values of $|t_{-1}|$ are preferred. 
The largest  values of $|t_{-1}|$ in Figure \ref{n3common}  are obtained for clusters oriented vertically. 
Interestingly,  the same figure indicates that zero transmission points occur at cluster angles perpendicular or parallel to the incident wavefront. 
For detailed investigation we select $(d,\theta_d)$ pairs with large values of $|t_{-1}|$, namely $(0.225,\pi/2)$ and $(0.751,\pi/2)$. The corresponding impedances of the scatterers are listed in table \ref{tabclusters}. Note that the  the two selected clusters differ only in the (vertical) spacing $d$, and that the difference between the two values, $d_2-d_1\approx 0.5$ corresponds to a phase change of $k_y (d_2-d_1) \approx \pi / 2$.  Other points with the same high transmission correspond to $2 \pi$ phase change in the $y$ direction, and are situated outside the region shown above (and below) clusters \circled{1} and \circled{2}.

It might seem surprising that the optimal orientation of the linear cluster is vertical, since it is clear from \eqref{27.1} 
and  the identities  
${\bf k}_{-1}^\pm = - {\bf k}_{0}^\mp$ for the negative refractor, 
that if the three scatterers are on a line parallel to the $y-$axis then the second and third rows of  $ \hat{\bf S}$ are identical, making the matrix singular. 
However, it is  shown in Appendix \ref{B} that even though 
 the matrix $ \hat{\bf S}$ is indeed singular for $\theta_d = \pi /2$, the  
vector $\hat{\bf S}^{-1} {\bf e}_1$ which appears in \eqref{735} remains finite.
The symmetry of the $3-$cluster for $\theta_d = \pi /2$ also implies that the matrix 
$\pmb \chi$ of \eqref{13} is symmetric with only three independent elements, since 
$\chi_{11} = \chi_{22}= \chi_{33}$  and $\chi_{12} = \chi_{13}$.

\begin{figure}[h!]
	\centering
	\includegraphics[width=0.78\linewidth]{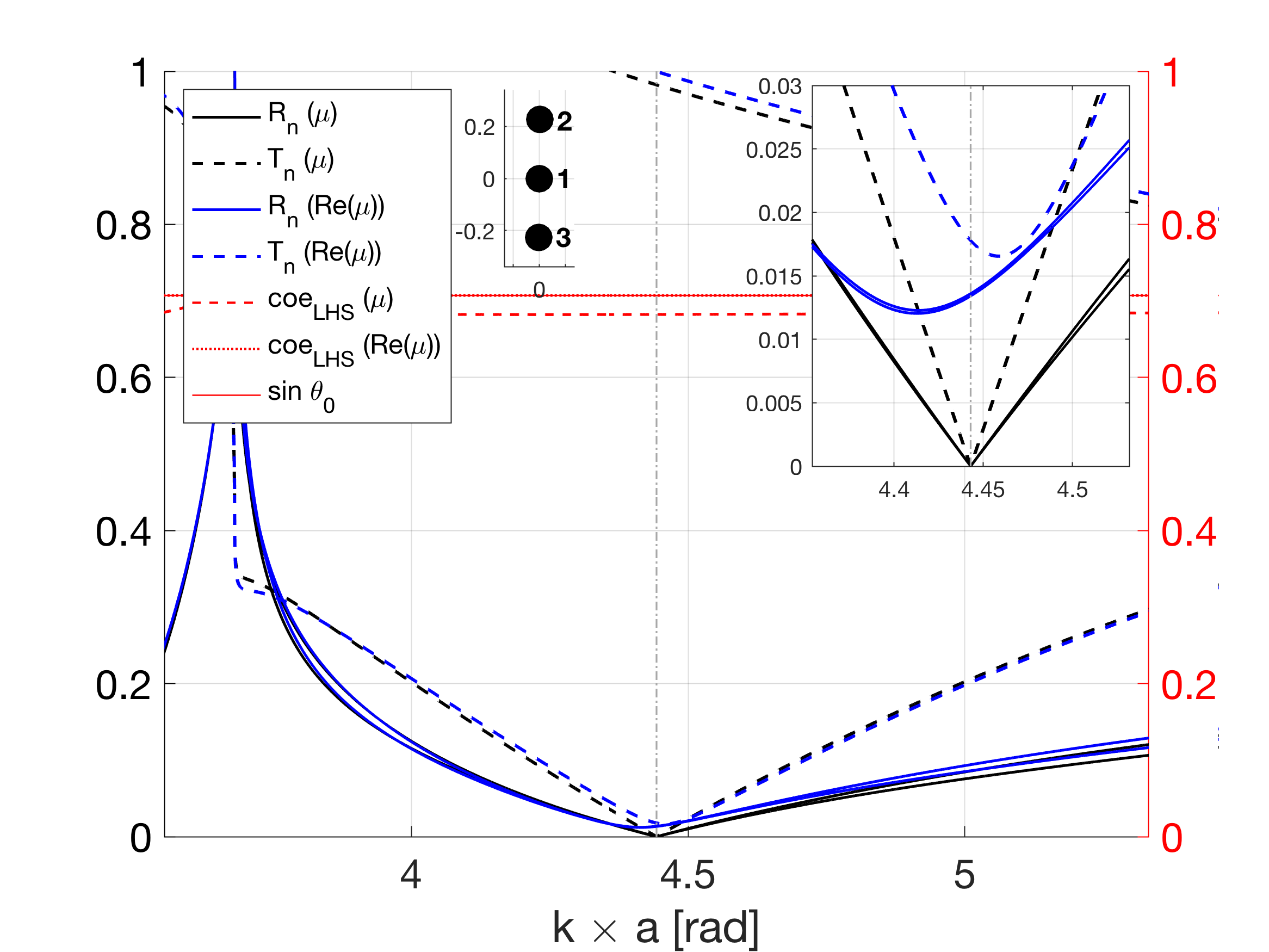}
	\caption{Reflection and transmission coefficients for the linear cluster \textcircled{1} computed for impedances given in table \ref{tabclusters} for complex values of $\mu_\alpha$ and for $\imag \mu_\alpha = 0$. The vertical dash-dot line at $k \times a = \pi / \cos \pi/4$ indicates the operating point of the grating.}
	\label{rtcoeffsline1}
\end{figure}

\begin{figure}[h!]
	\centering
	\includegraphics[width=0.78\linewidth]{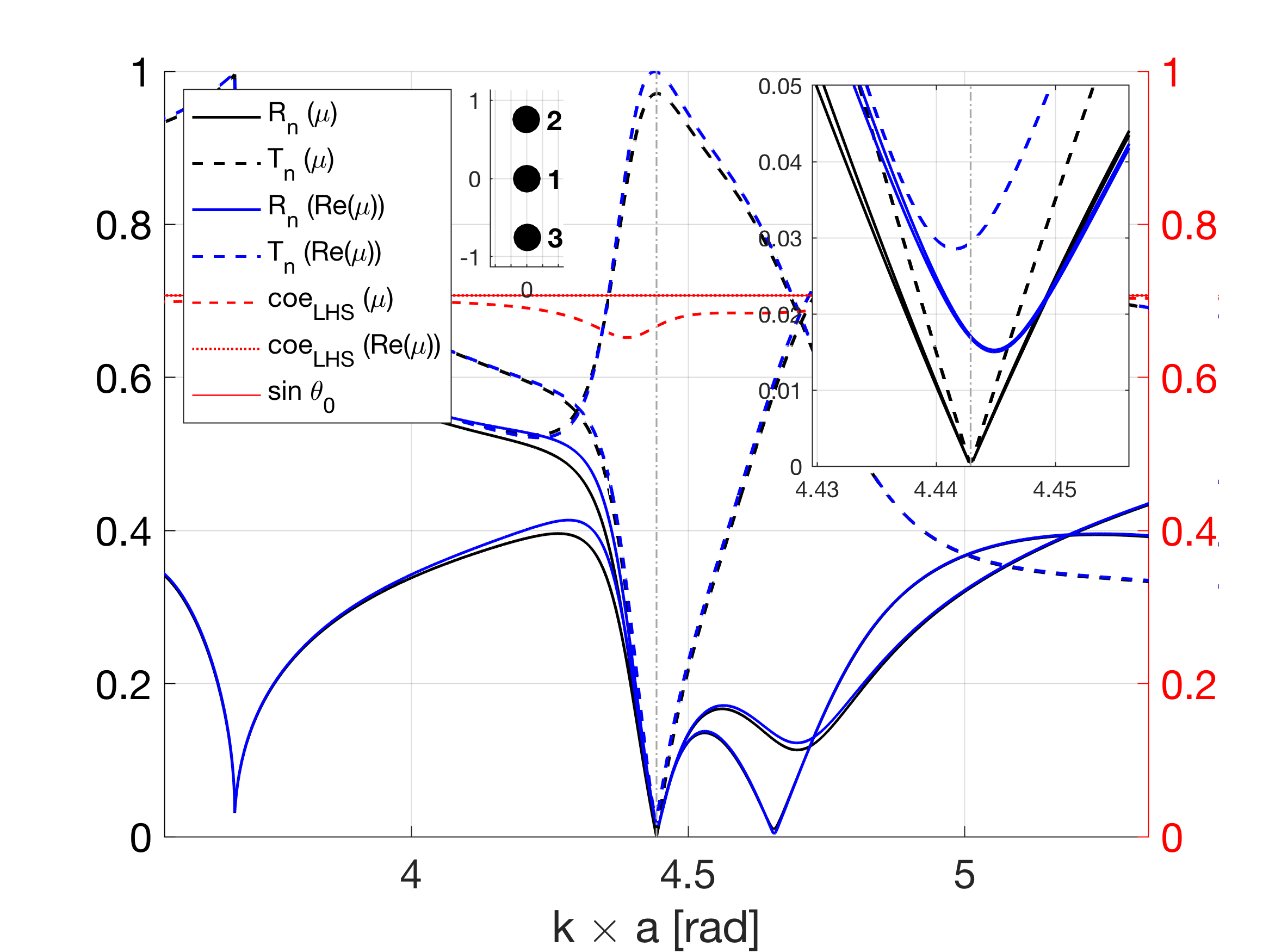}
	\caption{Reflection and transmission coefficients for the linear cluster \textcircled{2} using impedances from table \ref{tabclusters} for complex values of $\mu_\alpha$ and for $\imag \mu_\alpha = 0$. The vertical dash-dot line at $k \times a = \pi / \cos \pi/4$ indicates the operating frequency of the grating.}
	\label{rtcoeffsline2}
\end{figure}

Using impedances and cluster configurations from table \ref{tabclusters}, reflection and transmission coefficients, and plate displacements at the scatterers were computed for a wide range of $k \times a$. Results for the linear clusters \circled{1} and \circled{2} are shown in Figures \ref{rtcoeffsline1} and \ref{rtcoeffsline2}, respectively.

Brown solid horizontal lines in Figures \ref{rtcoeffsline1} and \ref{rtcoeffsline2} (and later) define the energy conservation threshold $\sin \theta_0$ of eq.\ \eqref{-33}, while the brown dotted lines depict the energy associated with all propagation modes, i.e.\ the left hand side (LHS) of eq.\ \eqref{-33}. Conservation of energy requires that  the continuous line is  above the dotted one, which  is always the case in the examples considered.  

Figures \ref{rtcoeffsline1} and \ref{rtcoeffsline2} illustrate    relatively high transmission coefficients (approximately $0.97$) for the  $n = -1$ diffracted mode for the linear clusters, meaning that almost all energy incident on the grating is converted to this mode. Of the two configurations, $\circled{1}$ is more broadband, i.e. it achieves similar transmission properties for a wider  range of $k \times a$.

\begin{table}[t]
	\caption{Selected cluster configurations for $N=3$, see Figure \ref{N3clusters}.}
	\label{tabclusters}
	\begin{tabular}{|c||c|c||c|c|}
		\hline
		Cluster: & linear $\circled{1}$ & linear $\circled{2}$ & triangular $\circled{1}$ & triangular $\circled{2}$  \\
		\hline
		\hline
		$(d, \theta_{d})$ & $(0.225, \pi/2)$ & $(0.751, \pi/2)$ & $(0.7484,0.9237)$ & $(1.438, 1.281)$ \\
		\hline
		\hline
		$\mu_1 a^2/D$ & $9.7253 + 0.2878 \ii$ & $-1.2684 + 0.0095 \ii$ & $-1.8861 + 0.0141 \ii$ & $-3.4707 + 0.1240 \ii$ \\
		\hline
		$\mu_2 a^2/D$ & $2.2272 + 0.0022 \ii$ & $-0.8477 + 0.0043 \ii$ & $-0.6507 + 0.5238 \ii$ & $-4.0771 + 0.0308 \ii$ \\
		\hline
		$\mu_3 a^2/D$ & $2.1609 + 0.0370 \ii$ & $-0.8308 + 0.0120 \ii$ & $1.5165 + 0.0588 \ii$ & $-0.4929 + 0.3752 \ii$ \\
		\hline
		& $\times 10^{2}$ & $\times 10^{2}$ & $\times 10^{2}$ & $\times 10^{2}$ \\
		\hline
	\end{tabular}
\end{table}

We next relax the restrictions on the impedances  given in table \ref{tabclusters} by using only their real parts, with the results  shown in Figures \ref{rtcoeffsline1} and \ref{rtcoeffsline2} for linear clusters \circled{1} and \circled{2}, respectively. It can be seen that for both clusters, the reflection and transmission coefficients of the diffracted modes that were  previously almost zero  are now slightly increased, however, the target $t_{-1}$ coefficient is still near unity ($0.999$). Also, cluster \circled{1} displays better broadband characteristics than \circled{2}, the latter being more sensitive to precise selection of $k$.  It is interesting to note that cluster \circled{2} has small damping to begin with.  Also, the real parts of the impedances in both clusters are all  positive (cluster  \circled{1}) or negative  (cluster  \circled{2}).

\subsubsection{Results for the triangular cluster}

\begin{figure}[h!]
	\centering
	\begin{subfigure}{}
		\includegraphics[width=0.48\linewidth]{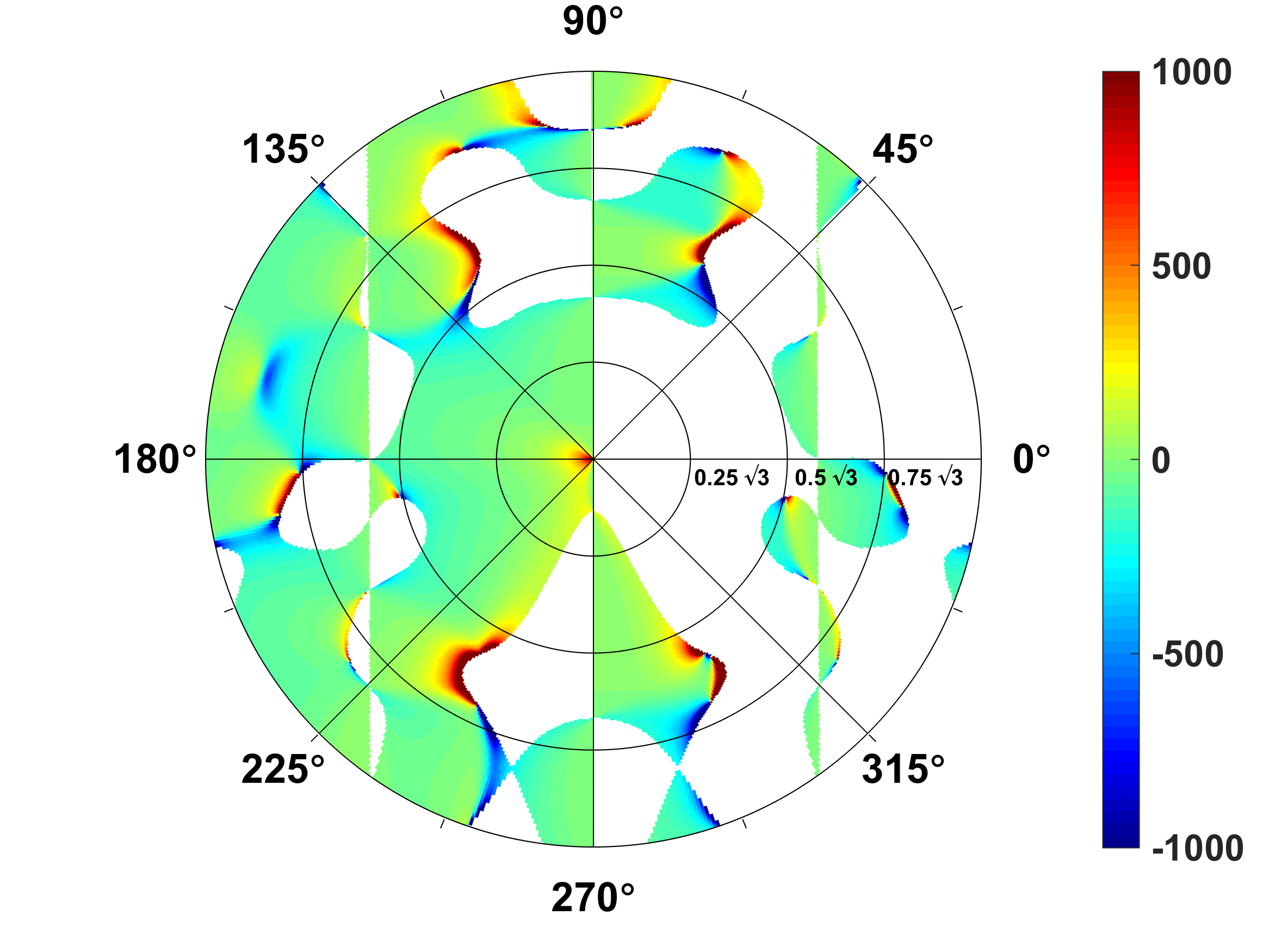}
	\end{subfigure}
	\begin{subfigure}{}
		\includegraphics[width=0.48\linewidth]{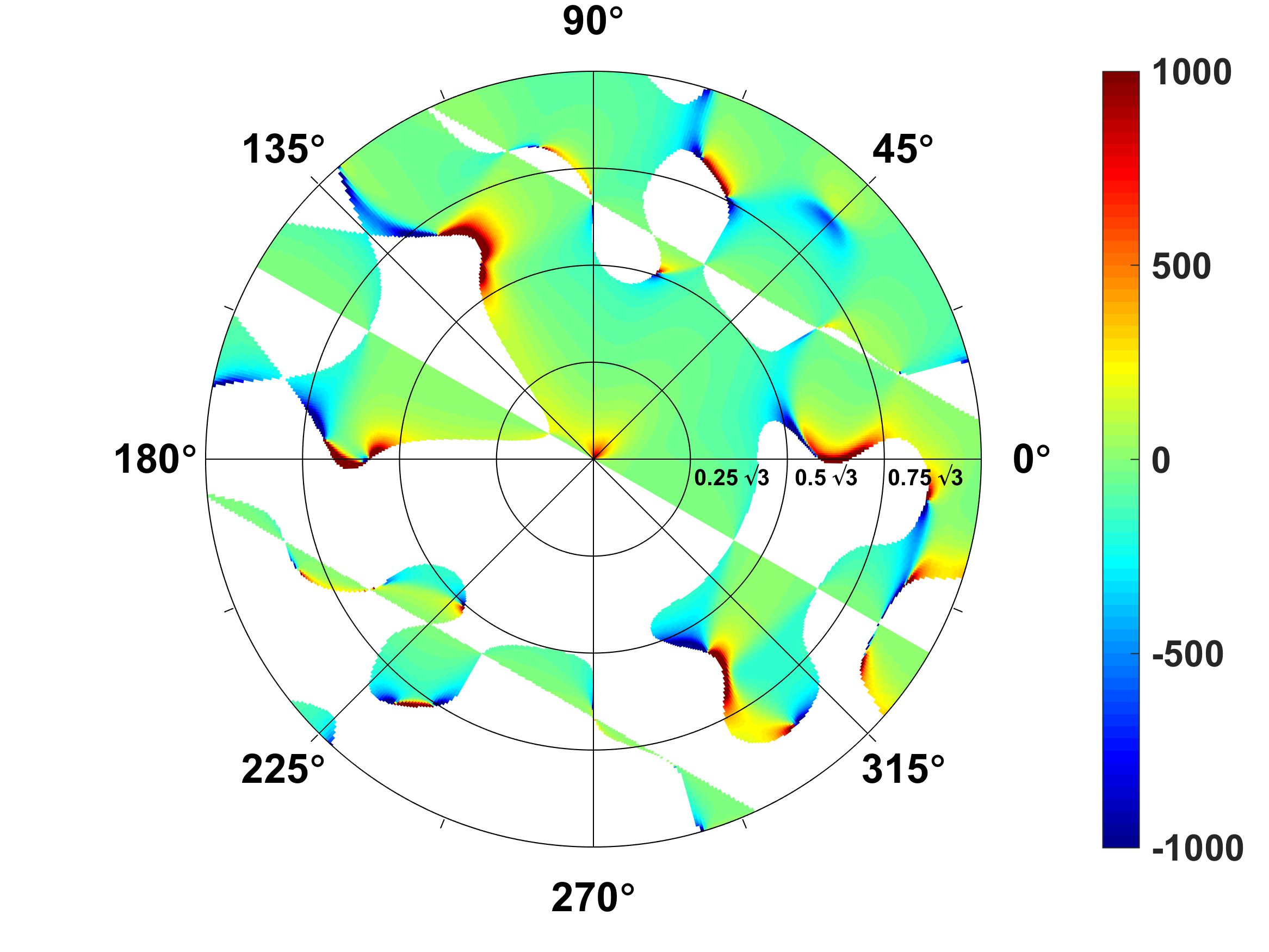}
	\end{subfigure}
	\caption{Real parts of the complex-valued impedance $\mu_1$ (left) and $\mu_2$ (right), in units of $a^2/D$, as functions of $d$ and $\theta_d$ for the incident wave  $\theta_0 = \pi/4$ and $k = \pi / (a \ \cos \theta_0)$ (with $a = 1$), for the $N=3$ triangular cluster negative refractor.}
	\label{n3mu1mu2mu3real}
\end{figure}

\begin{figure}[h!]
	\centering
	\begin{subfigure}{}
		\includegraphics[width=0.48\linewidth]{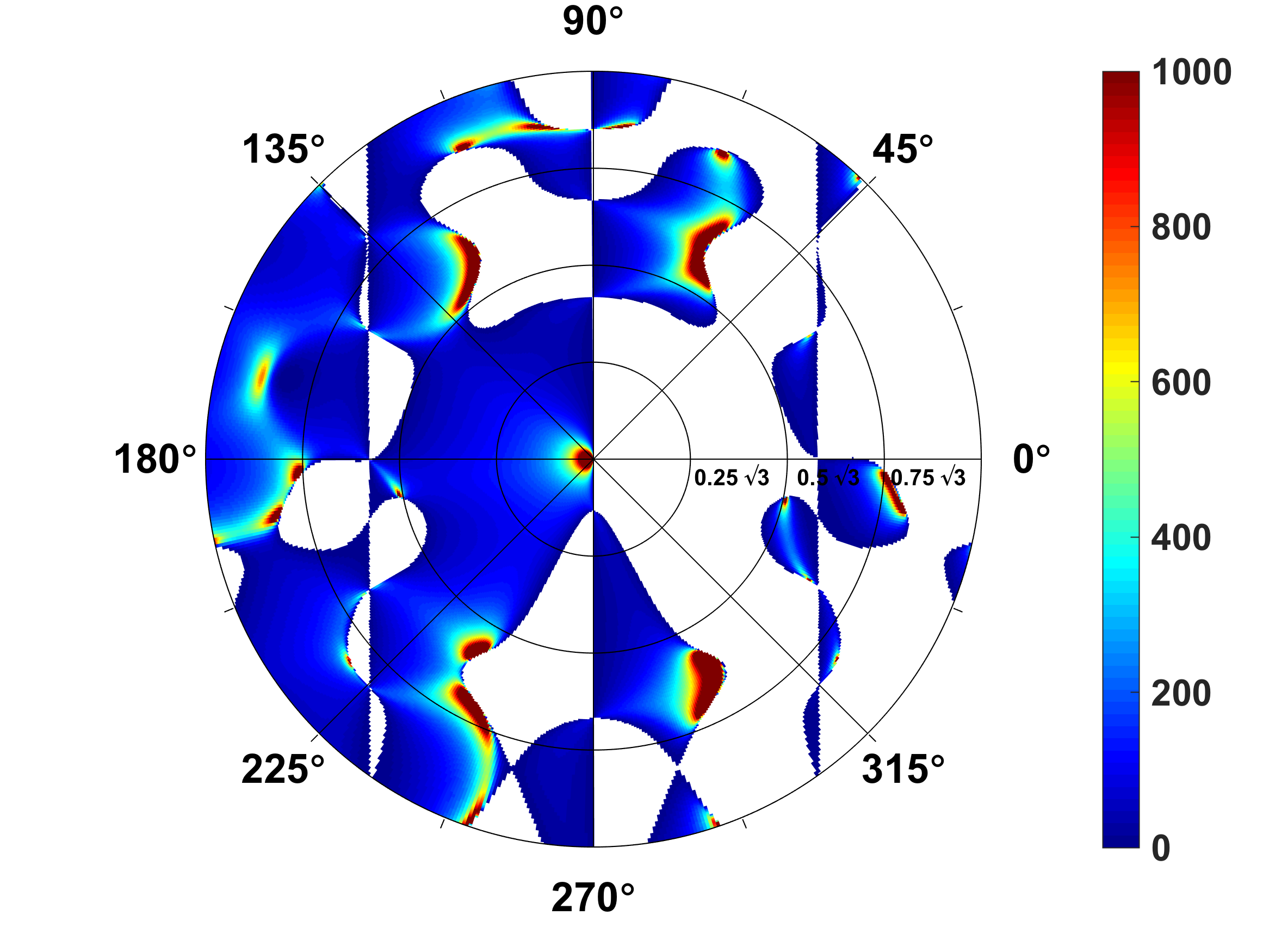}
	\end{subfigure}
	\begin{subfigure}{}
		\includegraphics[width=0.48\linewidth]{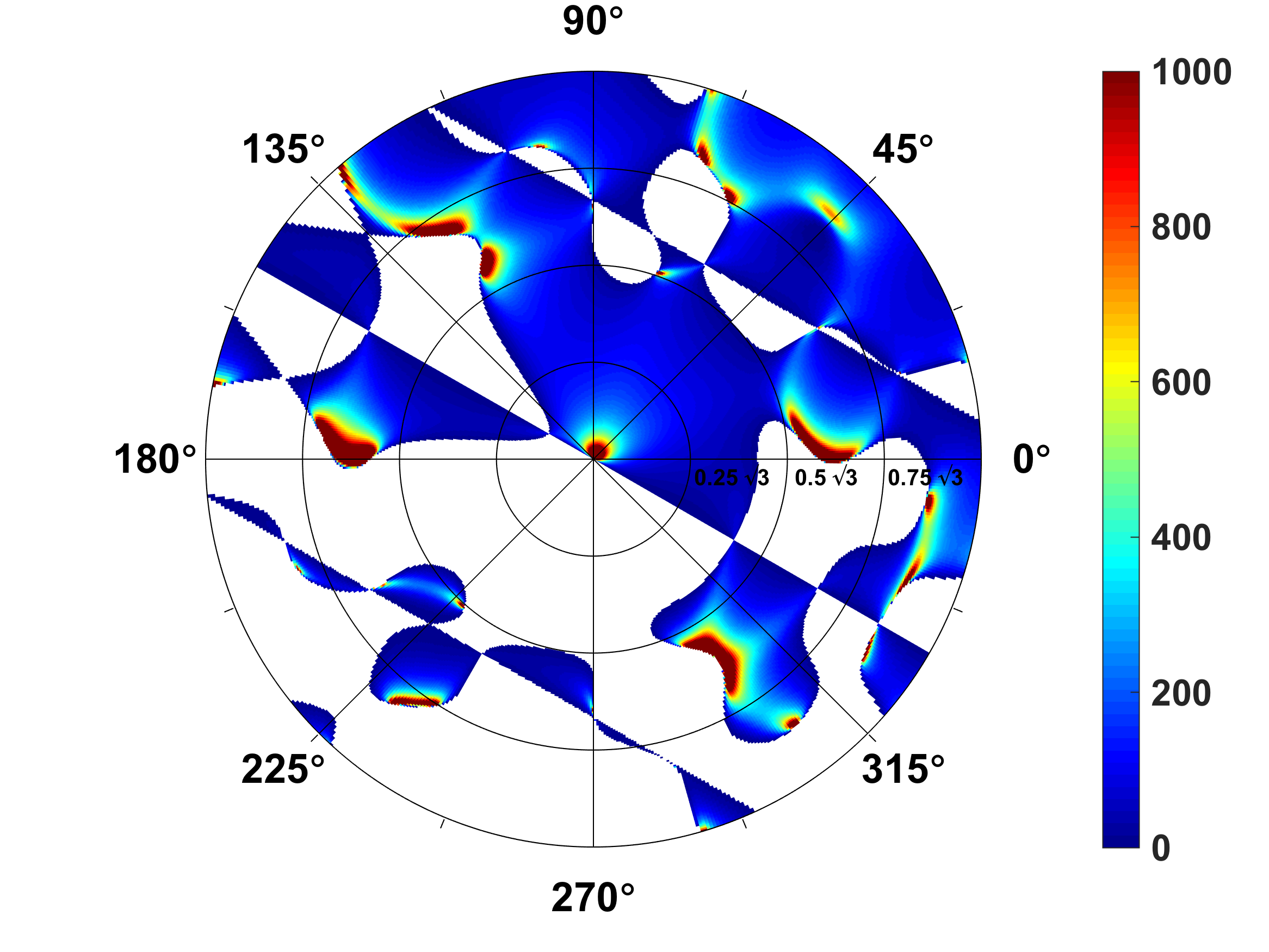}
	\end{subfigure}
	\caption{Imaginary parts of the  impedance $\mu_1$ (left) and $\mu_2$ (right), in units of $a^2/D$,  for the incident wave  $\theta_0 = \pi/4$, $k = \pi / (a \ \cos \theta_0)$ (for $a = 1$), for the $N=3$ triangular cluster negative refractor.}
	\label{n3mu1mu2mu3imag}
\end{figure}

\begin{figure}[h!]
	\centering
	\includegraphics[width=0.8\linewidth]{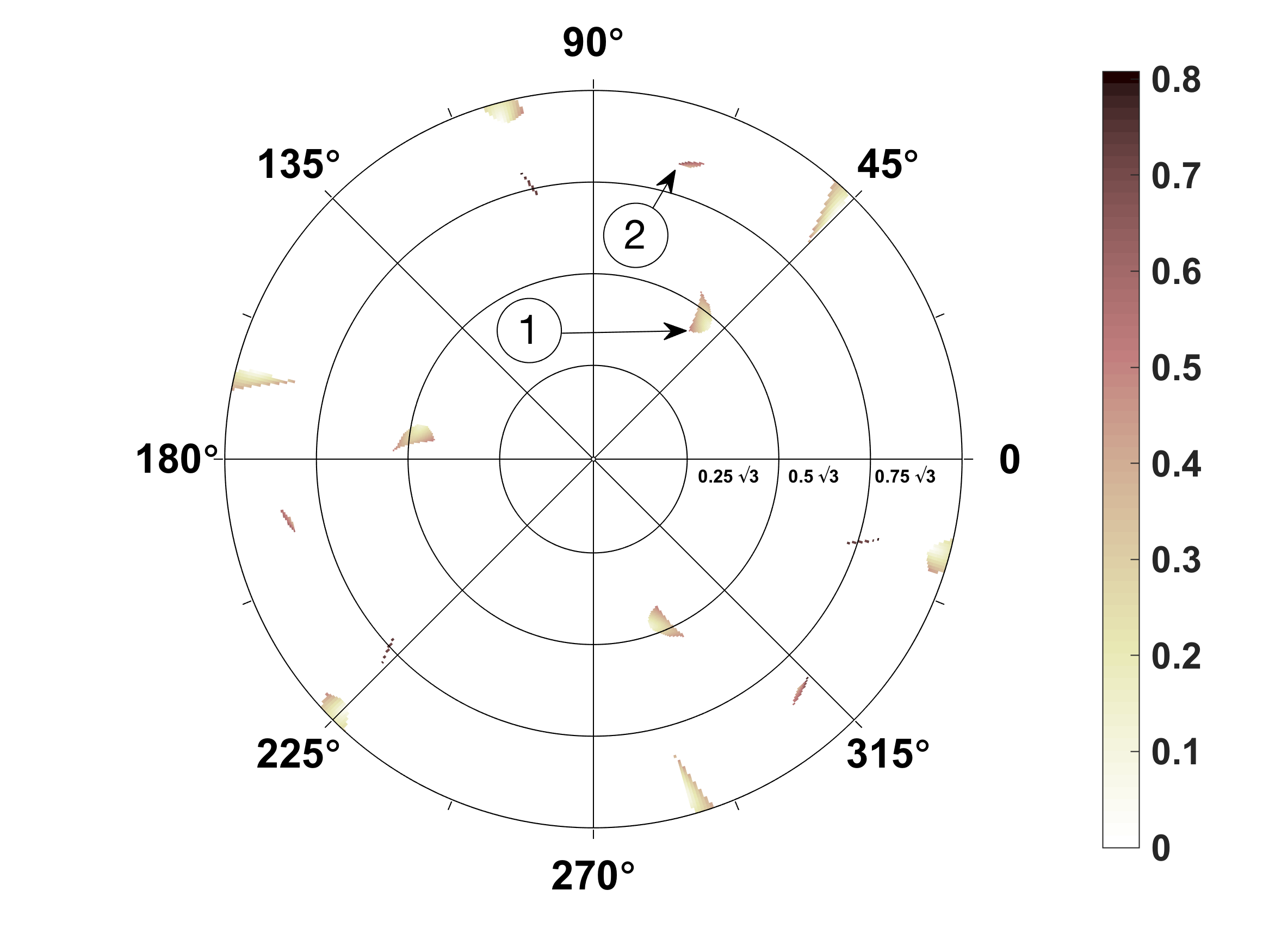}
	\caption{Transmission magnitude $|t_{-1}|$ of the designed triangular cluster 
negative refractor as a function of the scatterers' radial, $d$, and angular, $\theta_d$, coordinates.  
The plotted regions correspond to passive $\mu_\alpha$ for all $\alpha$, and the 
labels are the selected clusters 
of table \ref{tabclusters}.}
	\label{n3triacommon}
\end{figure}

Figures \ref{n3mu1mu2mu3real} and \ref{n3mu1mu2mu3imag} show, respectively, real and imaginary parts of the complex-valued impedance $\mu_1$ and (symmetric) $\mu_2$ ($\mu_3$ is also symmetric to $\mu_2$ due to the symmetry of the cluster) for $\theta_d \in (0, 2 \pi)$ and $d \in (0, a)$ for the triangular cluster. Again,  Figures \ref{n3mu1mu2mu3real} and \ref{n3mu1mu2mu3imag} only show  the parts of the $(d,\theta_d)$ plane for which $\imag \mu > 0$.
Combinations of $(d, \theta_d)$, with the values of $|t_{-1}|$ satisfying $\imag \mu_\alpha > 0$ for all $\alpha$, i.e. a passive cluster, are shown in Figure \ref{n3triacommon}.

\begin{figure}[h!]
	\centering
	\includegraphics[width=0.78\linewidth]{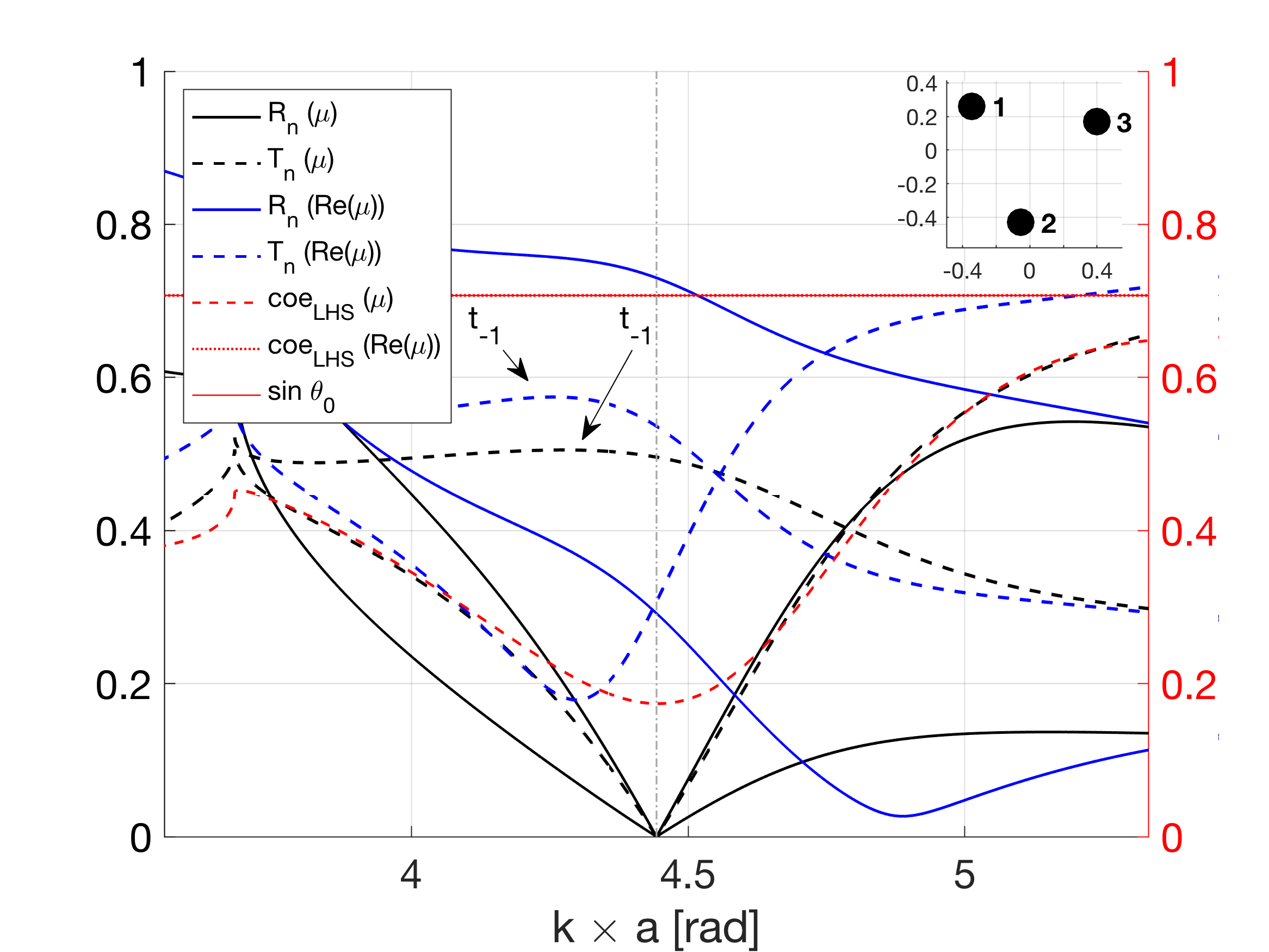}
	\caption{Reflection and transmission magnitudes for the triangular cluster \textcircled{1} defined in  table \ref{tabclusters} for complex values of $\mu_\alpha$ and for $\imag \mu_\alpha = 0$. The vertical dash-dot line at $k \times a = \pi / \cos \pi/4$ indicates the operating frequency.}
	\label{rtcoeffstriangle1}
\end{figure}

\begin{figure}[h!]
	\centering
	\includegraphics[width=0.78\linewidth]{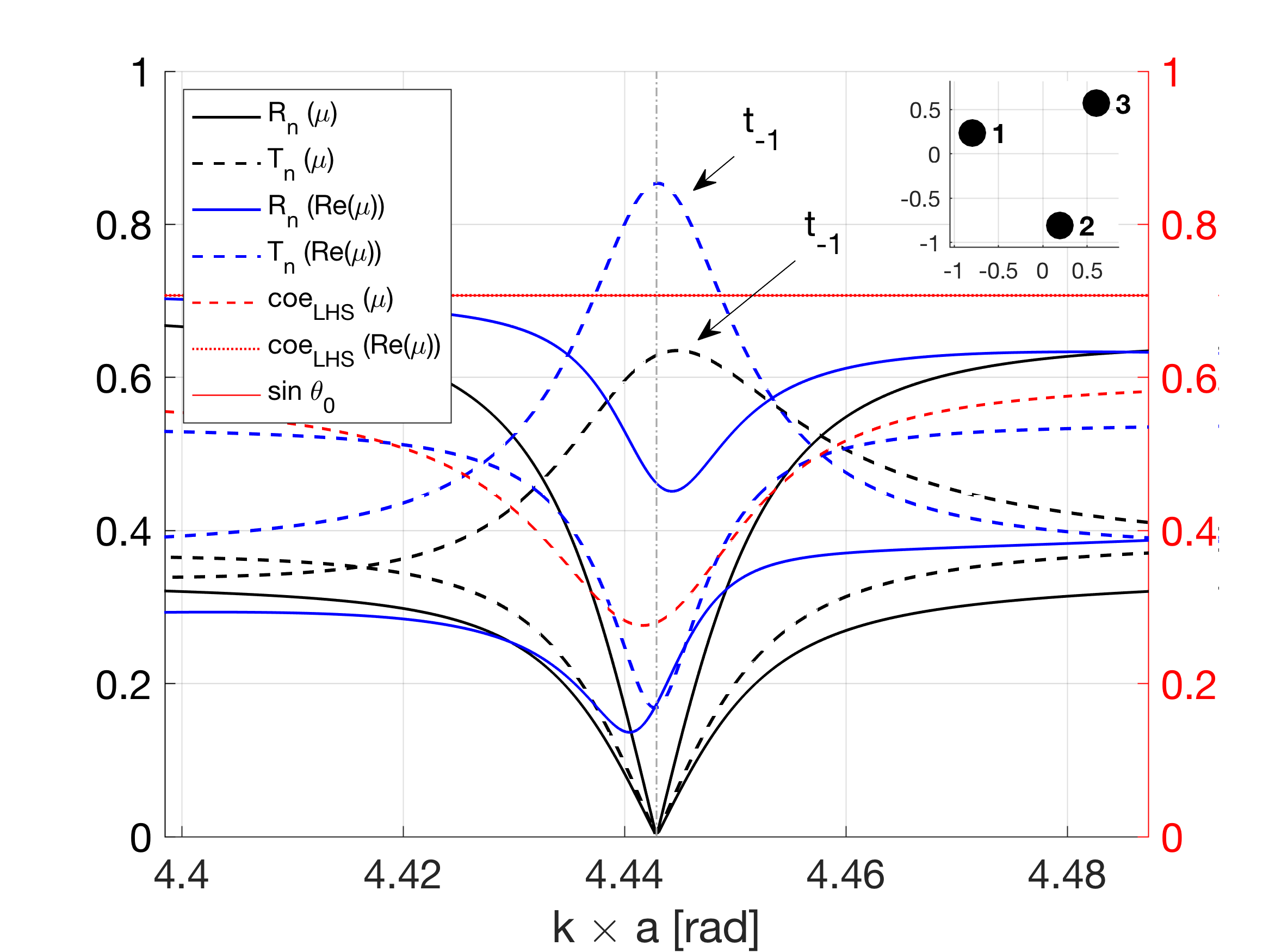}
	\caption{Reflection and transmission magnitudes for the triangular cluster \textcircled{2} with properties from  table \ref{tabclusters} for complex values of $\mu_\alpha$ and for $\imag \mu_\alpha = 0$. The vertical dash-dot line at $k \times a = \pi / \cos \pi/4$ identifies the operational frequency of the grating.}
	\label{rtcoeffstriangle2}
\end{figure}

As for the linear cluster, we select $(d,\theta_d)$ pairs with relatively large  values of $|t_{-1}|$. For the triangular cluster these are $(0.7484,0.9237)$ and $(1.438,1.281)$, with  impedances listed in table \ref{tabclusters}.
Figures \ref{rtcoeffstriangle1} and \ref{rtcoeffstriangle2}
show the reflection and transmission coefficients as a function of $k \times a$ for the chosen triangular clusters \circled{1} and \circled{2}. The triangular cluster \circled{1} displays moderate broadband response, while cluster \circled{2} is narrowband, thus sensitive to the frequency of the incident wave.

Figures \ref{rtcoeffstriangle1} and \ref{rtcoeffstriangle2} also show the reflection and transmission characteristics for the triangular clusters \circled{1} and \circled{2}, respectively, computed using only the real parts of the impedances, given in table \ref{tabclusters}.   Setting the imaginary parts of the impedances to zero results in  a significant drop in grating performance. The reflection and transmission coefficients that were zeroed out with the complex impedance now assume high values, exceeding the transmission coefficient of the $n = -1$ diffracted mode in all cases.   This contrasts with the linear clusters  for which the effect of setting   $\imag \mu_\alpha =0$ is minimal, see Figures \ref{rtcoeffsline1} and \ref{rtcoeffsline2}.   The difference can be explained by the observation from table \ref{tabclusters} that the impedances of the linear clusters are all lightly damped, while each of the triangular clusters has one impedance that is significantly damped.

\begin{figure}[h!]
	\centering
	\includegraphics[width=\linewidth]{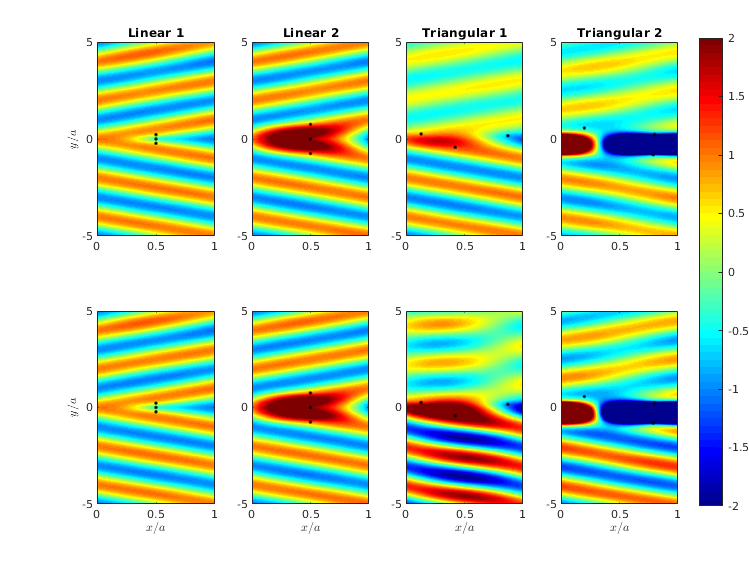}
	\caption{\label{fig:mst} Field maps of the diffraction of a plane wave by the different clusters of resonators as defined in Table \ref{tabclusters}. Upper panels show the full solution and lower panels show the same cluster but setting ${\imag( \mu_\alpha )= 0}$.  The black dots represent the point impedances within one period of the infinite grating. }
	\label{maps}
\end{figure}

\begin{figure}[h!]
	\centering
	\includegraphics[width=\linewidth]{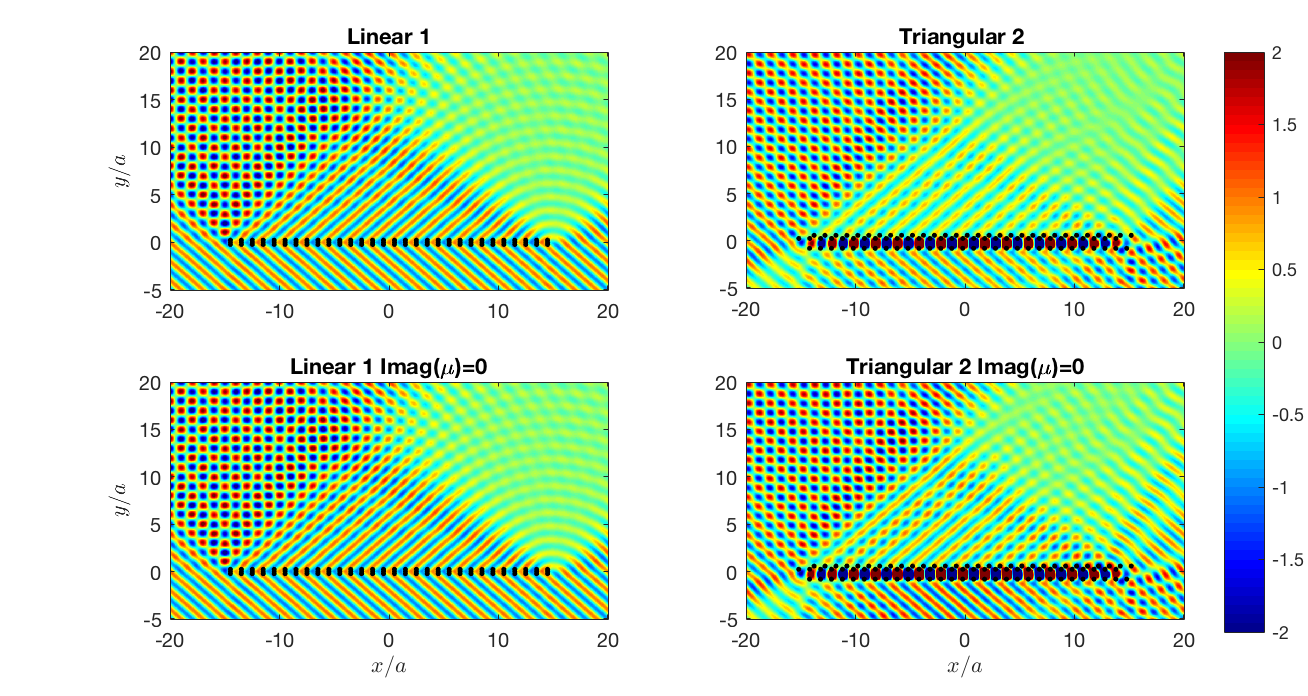}
	\caption{The total field amplitude for plane wave incidence on finite gratings of 30 clusters.  The parameters are otherwise the same as in Figure \ref{maps} for the infinite grating.} \label{fig:mst2} 
\end{figure}

\begin{figure}[h!]
	\centering
	\includegraphics[width=\linewidth]{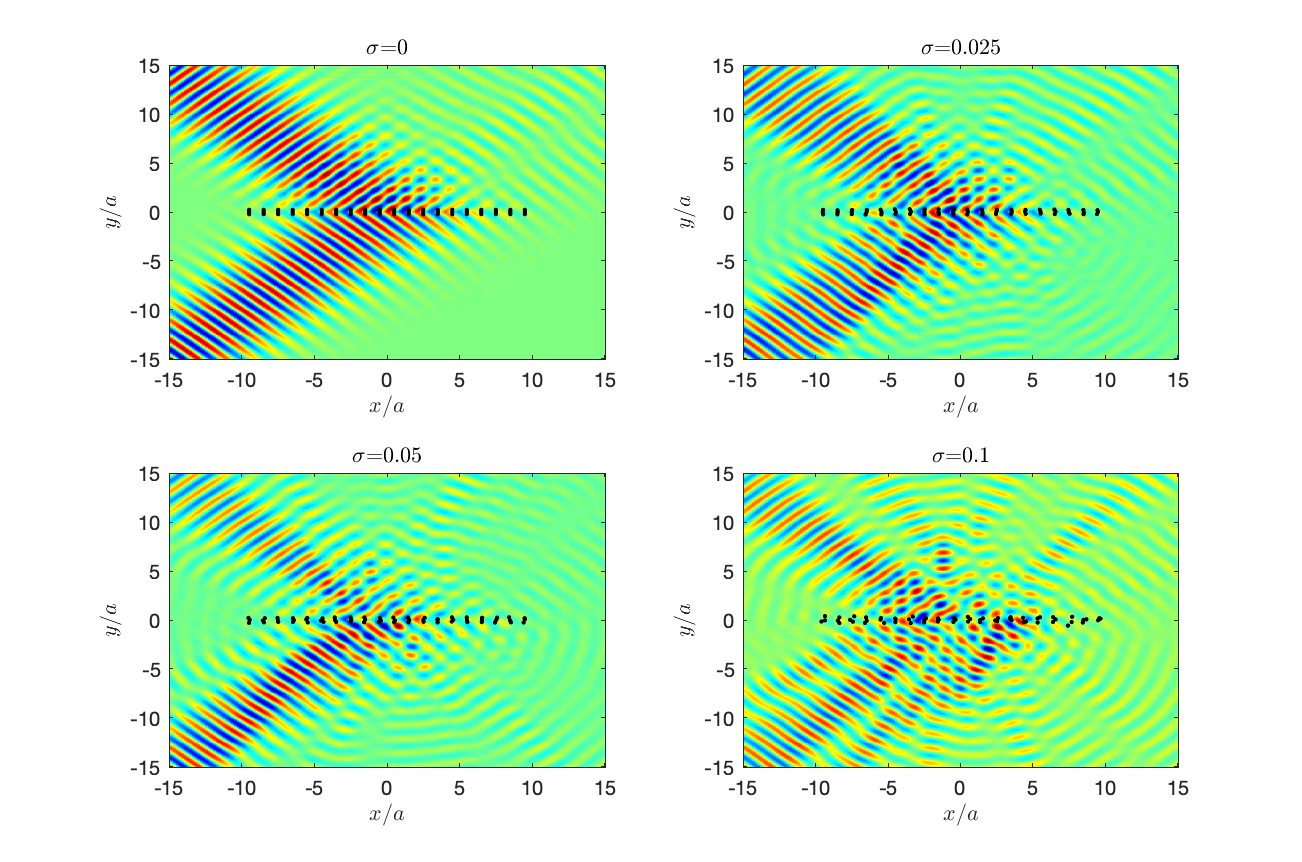}
	\caption{\rev{Multiple scattering simulations of finite gratings under Gaussian beam excitation. The parameter $\sigma$ represents the degree of random disorder that has been added to all the scatterers of the cluster.}} \label{fig:mst3} 
\end{figure}

\subsubsection{\rev{Infinite and finite retroreflector gratings with disorder}}

Figure \ref{fig:mst} shows the field distributions for the designed $N=3$ gratings. The upper panels show the simulation of an incident plane wave from the negative $y$ direction with incident angle $\theta_0 = \pi /4$ and wavenumber for the different configurations defined in Table \ref{tabclusters}. The lower panels show results for the same cluster after setting the imaginary part equal to zero. The negative refraction is evident in the simulations, and it is clear as well that, the larger the imaginary part of $\mu_\alpha$ the weaker the refracted wave.  This is a consequence of the loss of wave energy caused by  the highly damped resonators, although it is noted that the channeling of all the energy towards the $n=-1$ mode is still efficient in the sense that other modes are zeroed out, as designed. Overall, we see how ignoring the imaginary part has no visible effect in the linear cluster but  drastically diminishes the  amplitude of the refracted mode in the triangular clusters.  As noted above, the reason for this may be understood  from  the fact that scatterers of the linear clusters are lightly damped but the triangular clusters have at least one highly damped impedance, see Table \ref{tabclusters}. 

\rev{Finally, Figures \ref{fig:mst2} and \ref{fig:mst3} demonstrate that the effects predicted for the infinite grating are robust under finite limitations on the grating size, for finite incident beams, and in the presence of positional disorder.  Thus, the same effects as observed for the infinite grating in Figure \ref{maps} are apparent in Figure \ref{fig:mst2} which shows the total field for incidence on a finite grating of 30 clusters of the linear and triangular configurations.  The same finite configuration is considered in Figure \ref{fig:mst3} for Gaussian beam incidence, and for imperfections in the grating.  The simulations indicate that good agreement with the infinite system under plane wave incidence is  expected for zero and small levels of disorder.   
}

\section{Practical considerations on scatterers and clusters}

\rev{The grating performance, in terms of its reflection and transmission properties, depends on deviations of actual operation conditions from designed ones. Here we consider  the performance as a function of deviations in    scatterer positions,   impedances,  and the operating wavelength. }

\rev{Anomalous refractors and reflectors are obviously narrowband, since the  effect is due to diffraction which  by definition is wavelength-dependent by  (see equation \eqref{-34}). However, this dependence is smooth, so that small deviations from the incident angle or desired wavelength  produce  small deviations in the diffracted angle. This is also true for the channeling of energy; as can be seen in Figures \ref{rtcoeffsline1} and \ref{rtcoeffsline2}, the frequency dependence of the energy exchange between modes is smooth around the optimal value. Small variations about  the optimal point  produce small additional scattered waves, while  the overall  effect remains unchanged.}

\begin{figure}[h]
	\centering
	\begin{subfigure}{\label{sens1a}}
		\includegraphics[width=0.48\linewidth]{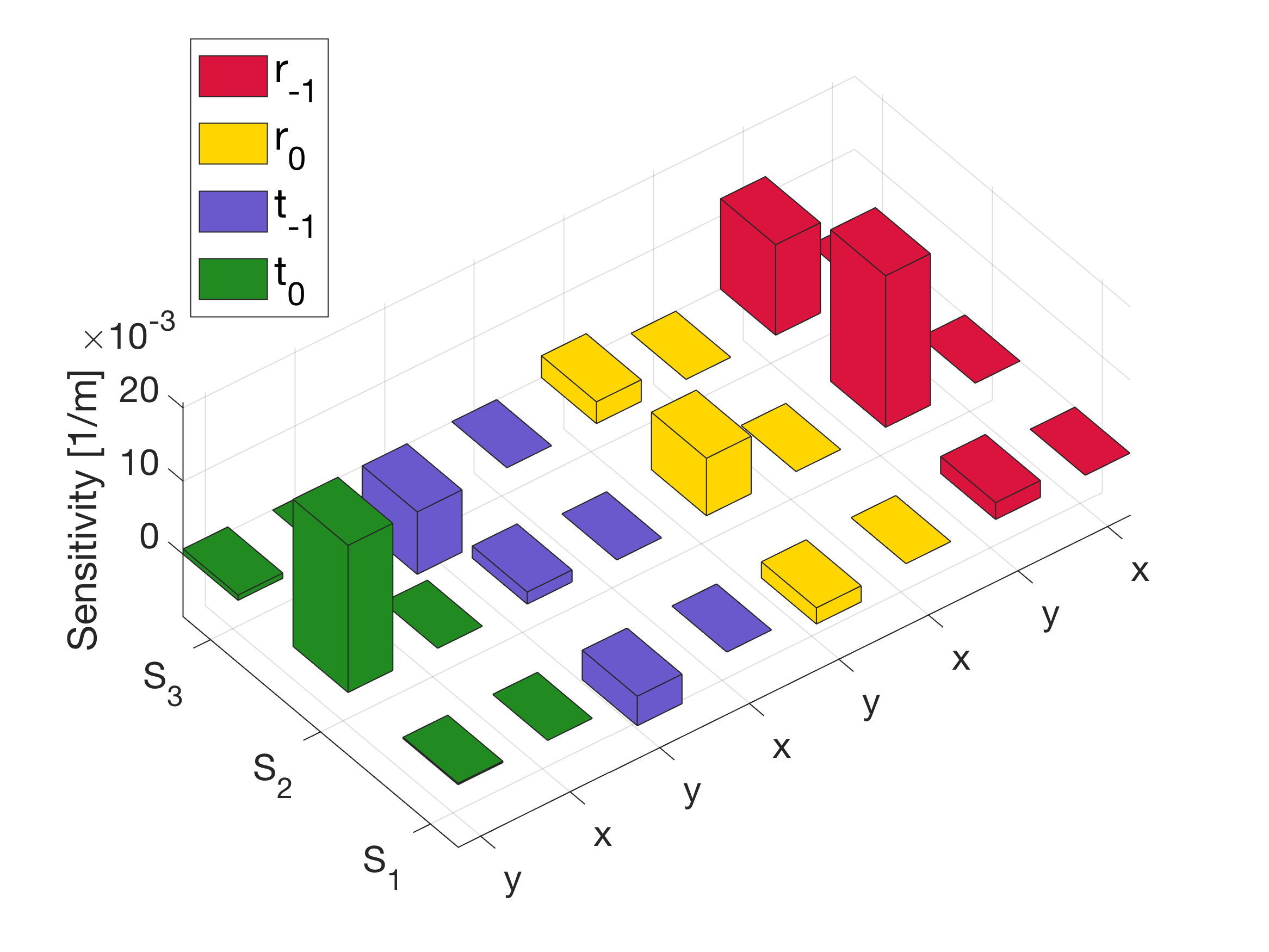}
	\end{subfigure}
	\begin{subfigure}{\label{sens1b}}
		\includegraphics[width=0.48\linewidth]{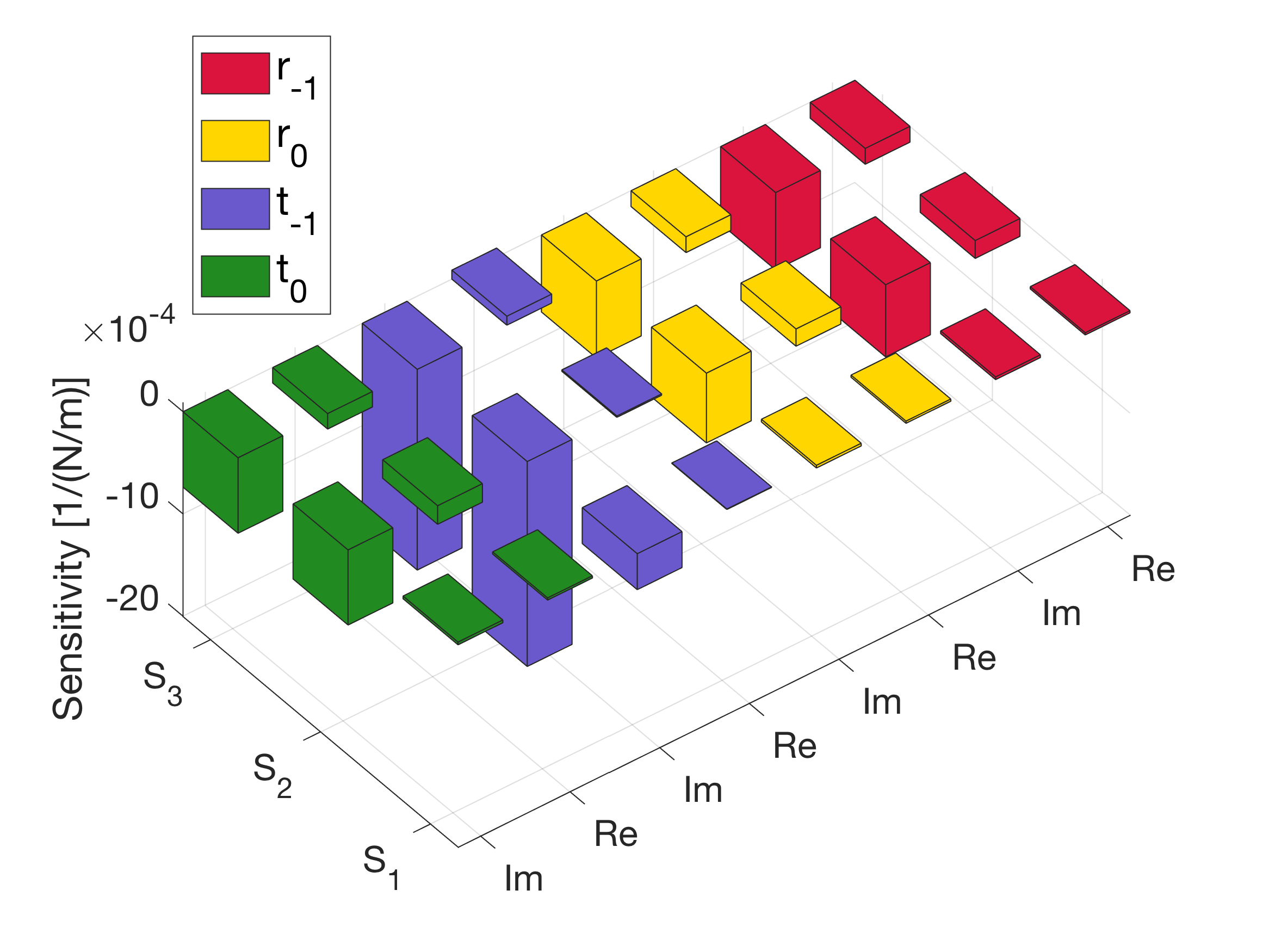}
	\end{subfigure}
	\caption{\rev{Sensitivity parameters for the linear \textcircled{1} cluster ($d = 0.225$, $\theta_d = \pi / 2$). (a) sensitivity parameters for individual changes in scatterers ($S_1$ - $S_3$) positions ($x$ and $y$) for the four diffracted modes reflection and transmission coefficients (for fixed $\mu$); (b) sensitivity parameters for individual changes in scatterers real and imaginary parts of impedances ($\mu_1$ - $\mu_3$) for the four diffracted modes reflection and transmission coefficients (fixed $S_1$ - $S_3$ positions).}}
	\label{sens1}
\end{figure}

\begin{figure}[h]
	\centering
	\begin{subfigure}{\label{sens2a}}
		\includegraphics[width=0.48\linewidth]{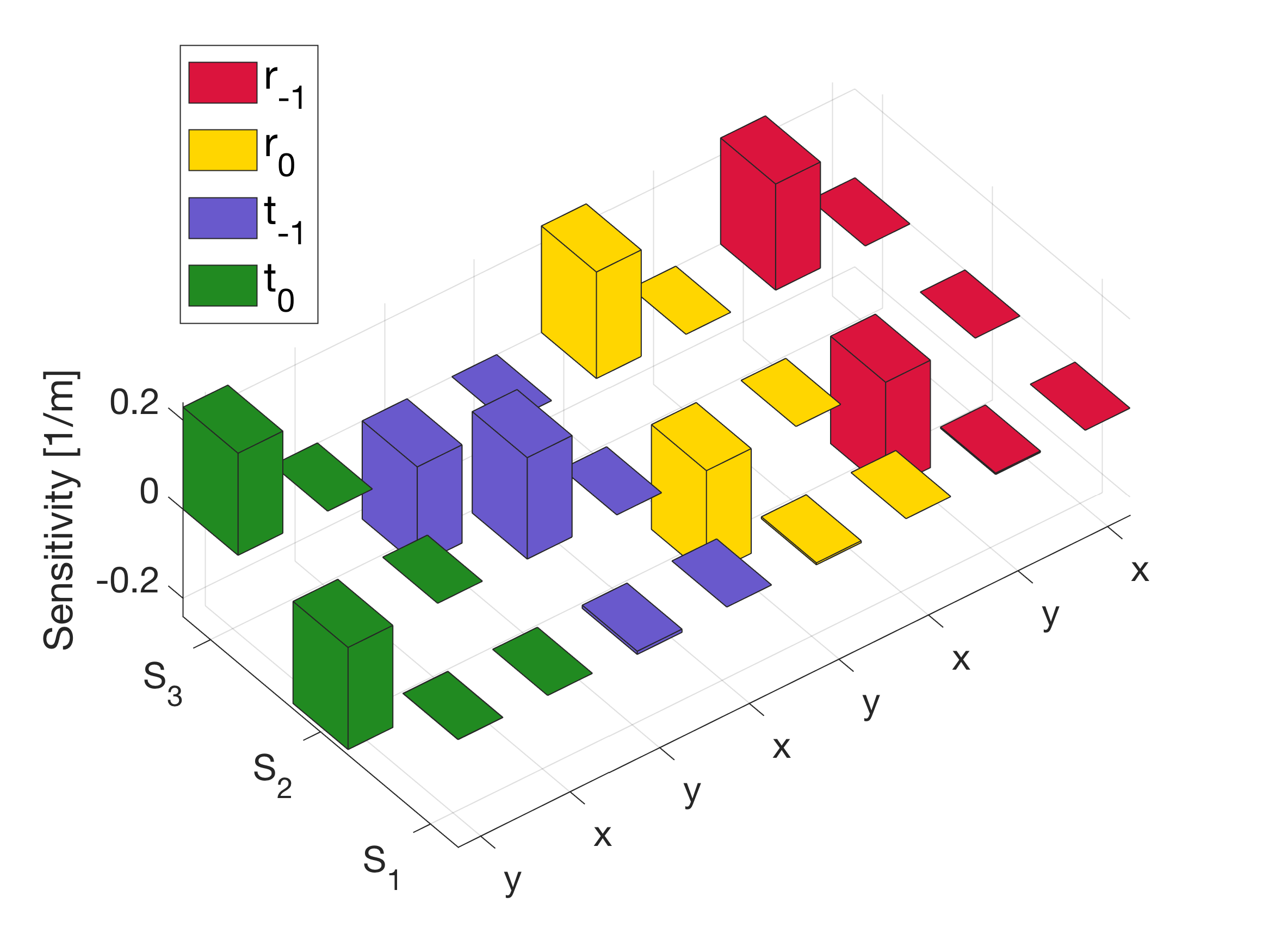}
	\end{subfigure}
	\begin{subfigure}{\label{sens2b}}
		\includegraphics[width=0.48\linewidth]{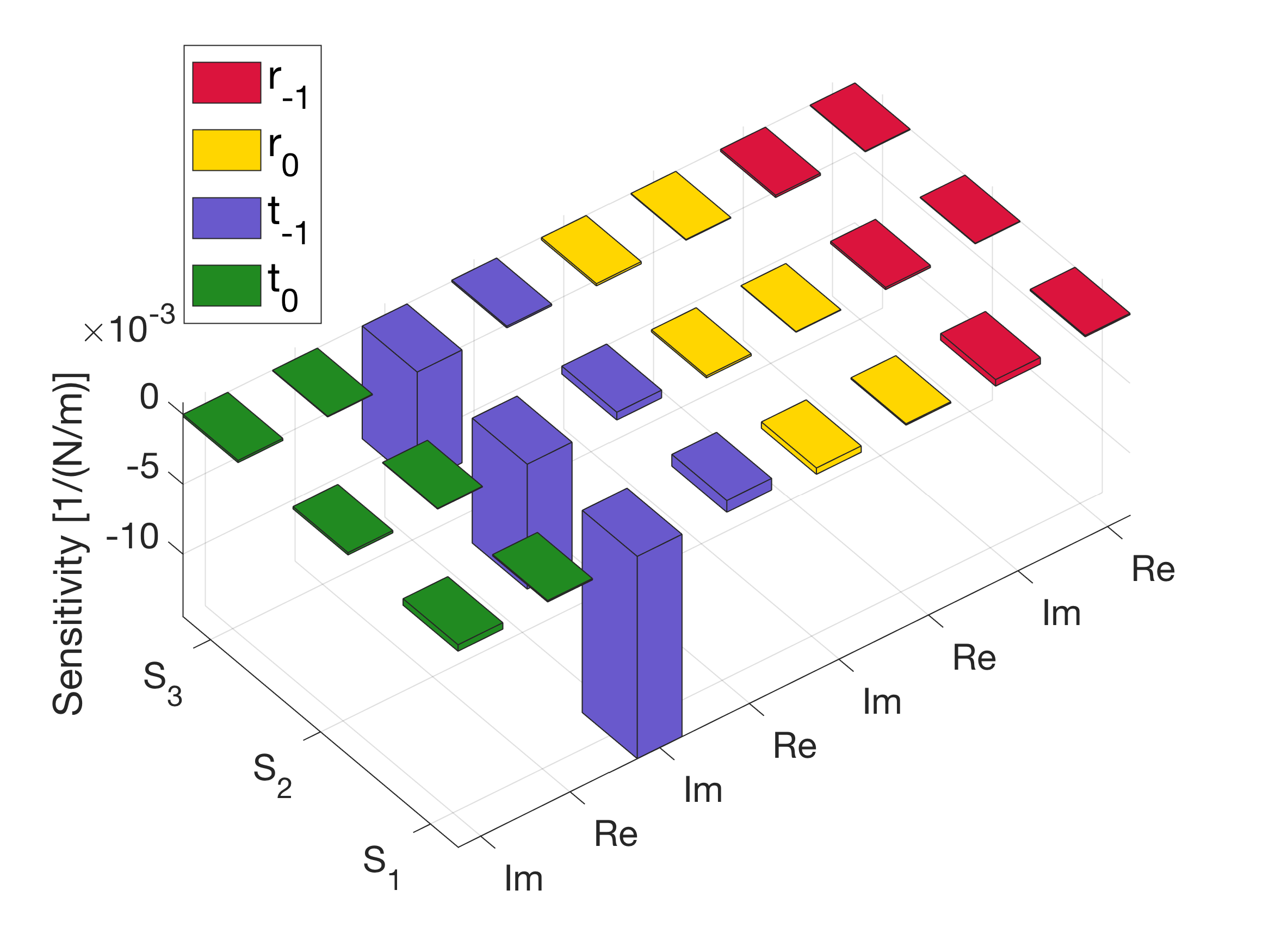}
	\end{subfigure}
	\caption{\rev{Sensitivity parameters for the linear \textcircled{2} cluster ($d = 0.75$, $\theta_d = \pi / 2$). (a) sensitivity parameters for individual changes in scatterers ($S_1$ - $S_3$) positions ($x$ and $y$) for the four diffracted modes reflection and transmission coefficients (for fixed $\mu$); (b) sensitivity parameters for individual changes in scatterers real and imaginary parts of impedances ($\mu_1$ - $\mu_3$) for the four diffracted modes reflection and transmission coefficients (fixed $S_1$ - $S_3$ positions).}}
	\label{sens2}
\end{figure}

\begin{figure}[h]
	\centering
	\begin{subfigure}{\label{sens3a}}
		\includegraphics[width=0.48\linewidth]{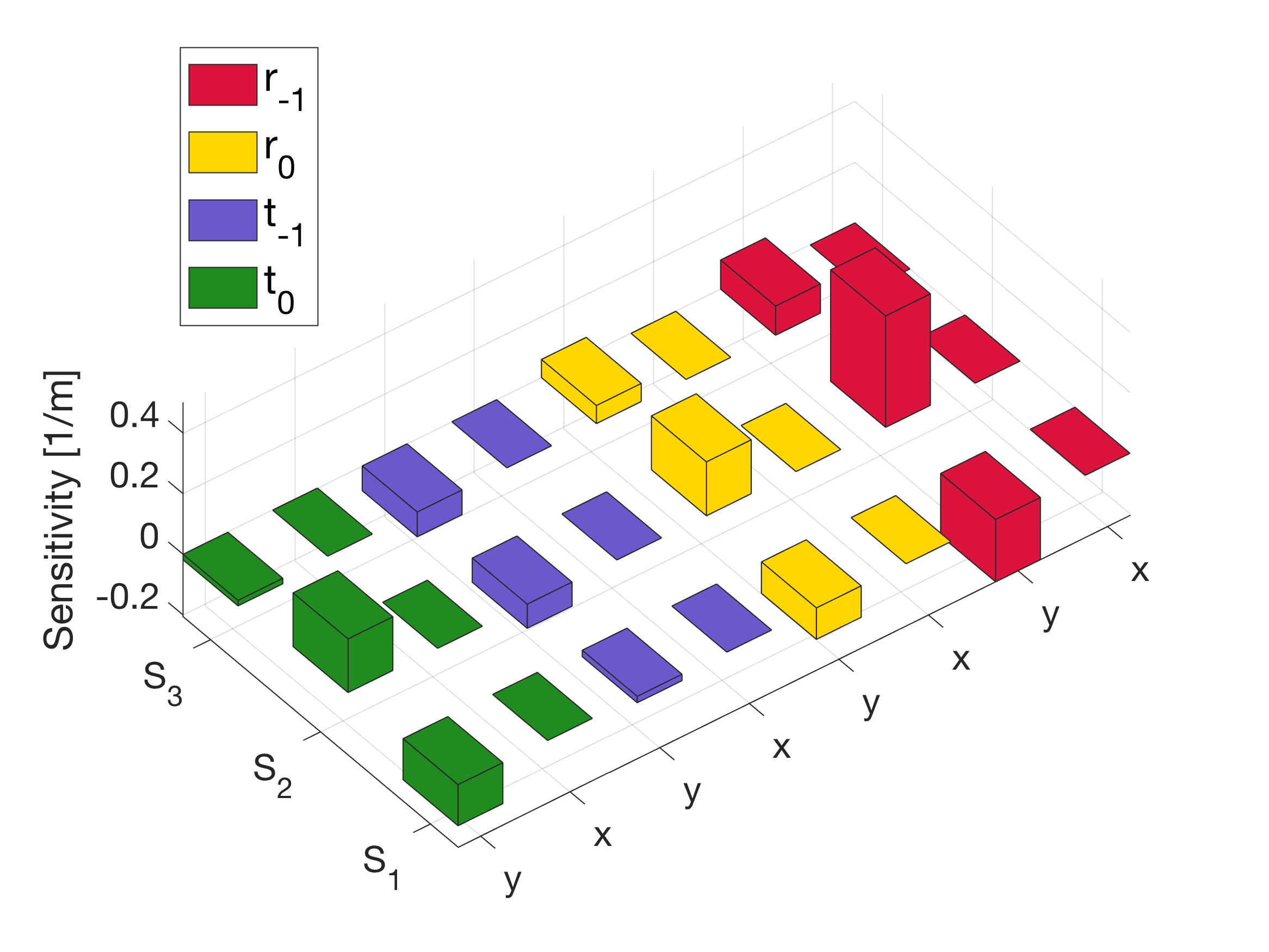}
	\end{subfigure}
	\begin{subfigure}{\label{sens3b}}
		\includegraphics[width=0.48\linewidth]{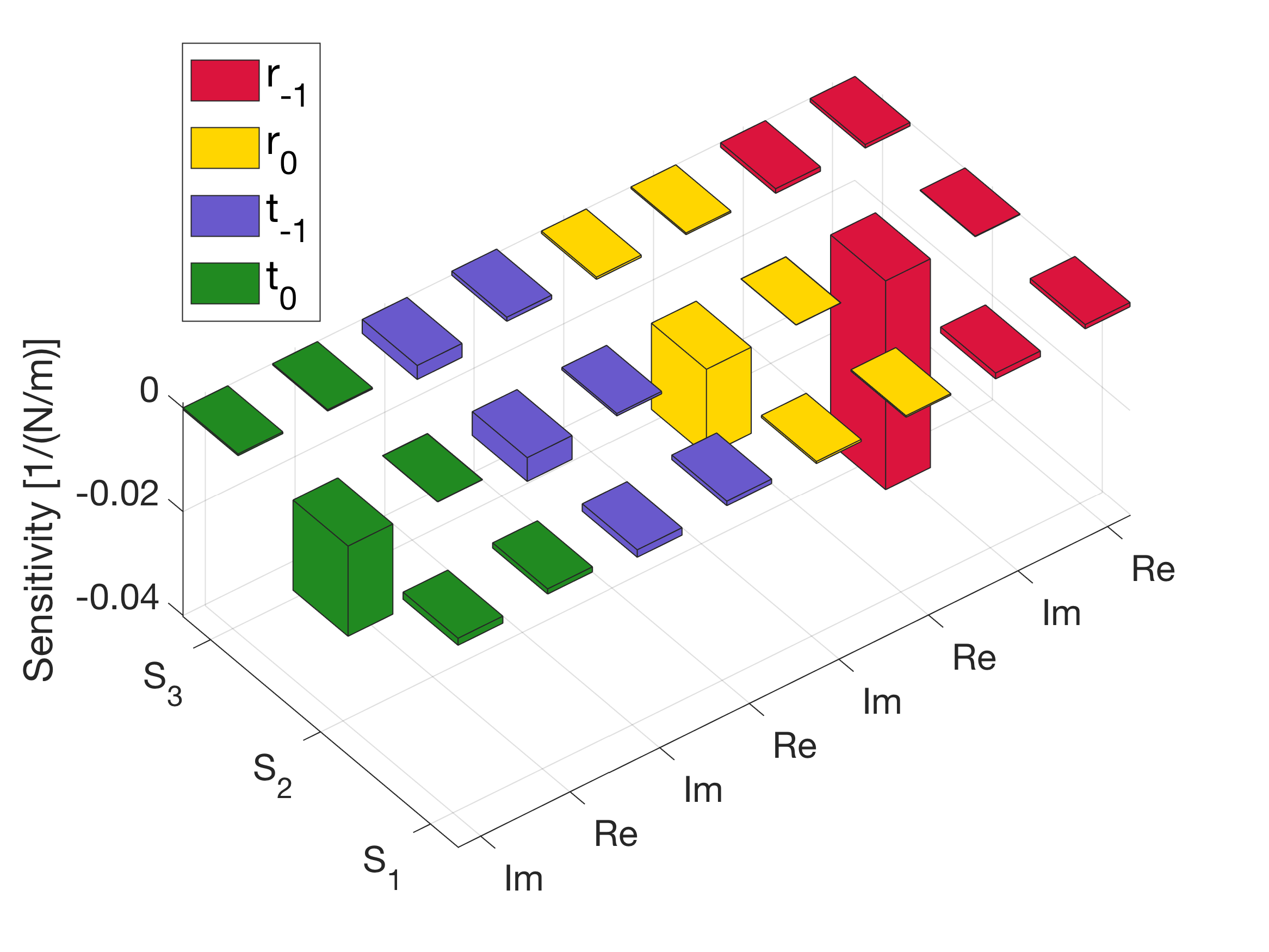}
	\end{subfigure}
	\caption{\rev{Sensitivity parameters for the triangular \textcircled{1} cluster ($d = 0.75$, $\theta_d = 0.92 $). (a) sensitivity parameters for individual changes in scatterers ($S_1$ - $S_3$) positions ($x$ and $y$) for the four diffracted modes reflection and transmission coefficients (for fixed $\mu$); (b) sensitivity parameters for individual changes in scatterers real and imaginary parts of impedances ($\mu_1$ - $\mu_3$) for the four diffracted modes reflection and transmission coefficients (fixed $S_1$ - $S_3$ positions).}}
	\label{sens3}
\end{figure}

\begin{figure}[h]
	\centering
	\begin{subfigure}{\label{sens4a}}
		\includegraphics[width=0.48\linewidth]{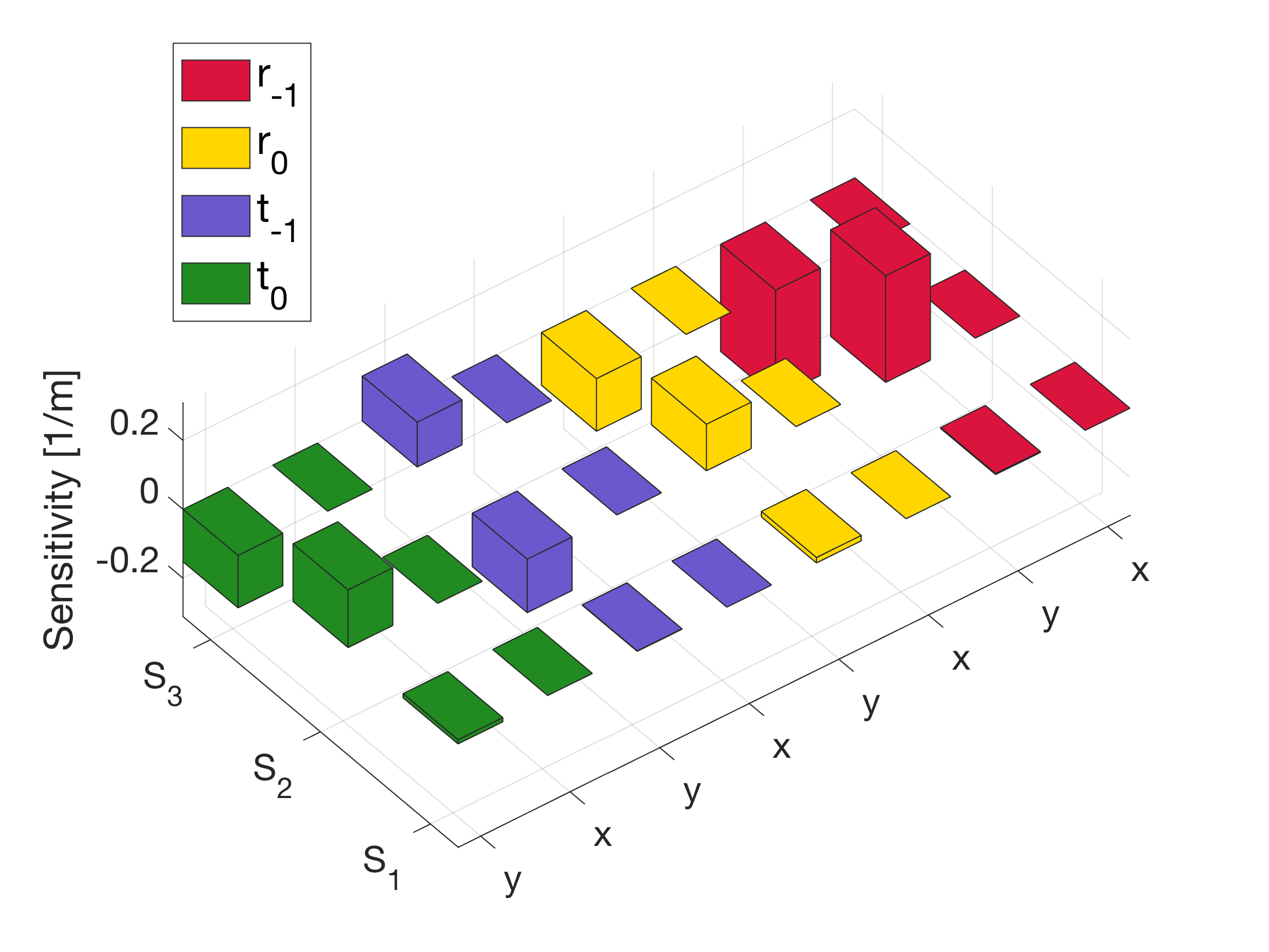}
	\end{subfigure}
	\begin{subfigure}{\label{sens4b}}
		\includegraphics[width=0.48\linewidth]{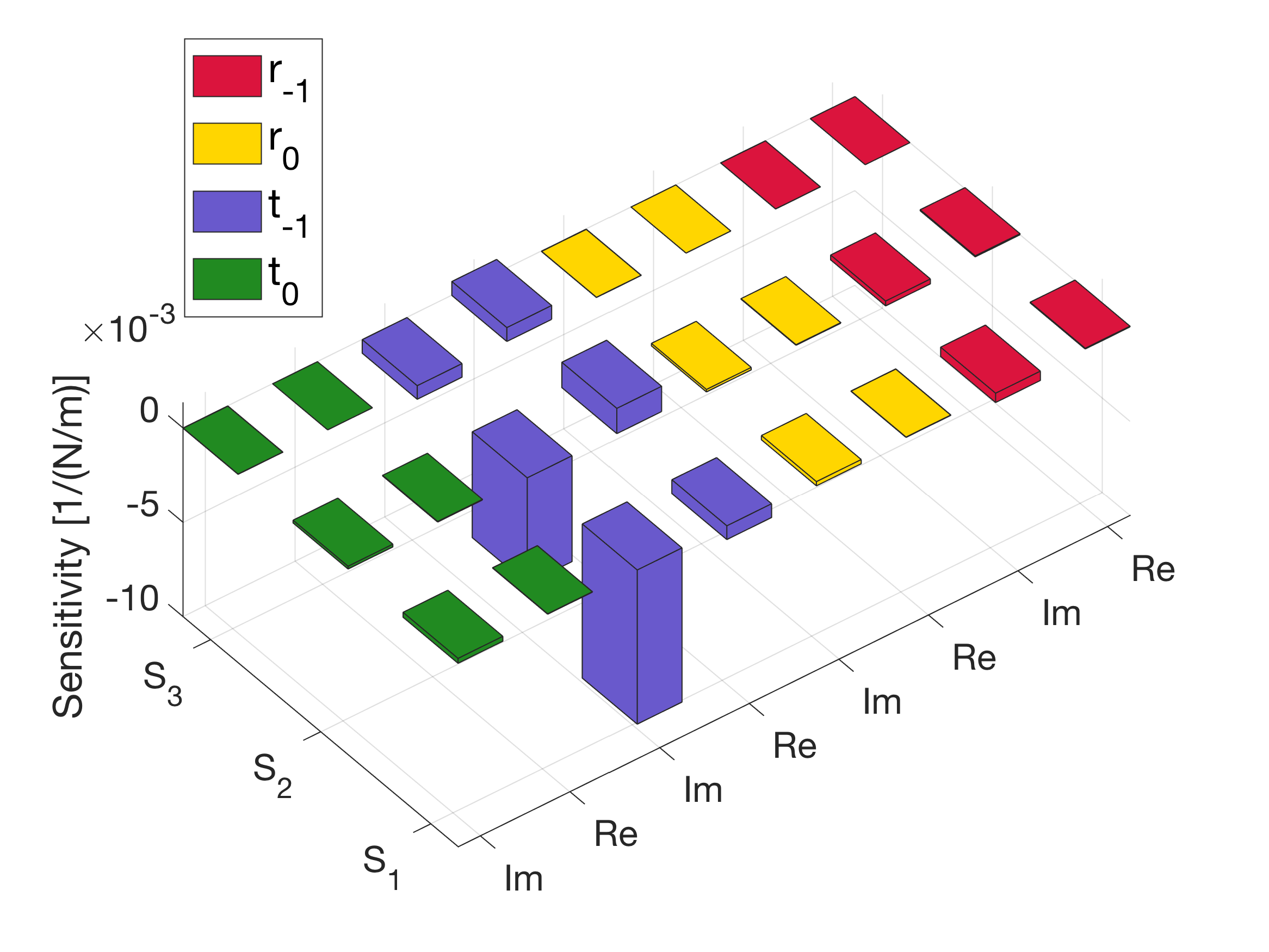}
	\end{subfigure}
	\caption{\rev{Sensitivity parameters for the triangular \textcircled{2} cluster ($d = 0.75$, $\theta_d = 0.92 $). (a) sensitivity parameters for individual changes in scatterers ($S_1$ - $S_3$) positions ($x$ and $y$) for the four diffracted modes reflection and transmission coefficients (for fixed $\mu$); (b) sensitivity parameters for individual changes in scatterers real and imaginary parts of impedances ($\mu_1$ - $\mu_3$) for the four diffracted modes reflection and transmission coefficients (fixed $S_1$ - $S_3$ positions).}}
	\label{sens4}
\end{figure}

\rev{We have already seen in in Figures \ref{rtcoeffsline1}, \ref{rtcoeffsline2}, \ref{rtcoeffstriangle1} and \ref{rtcoeffstriangle2} how the reflection and transmission coefficients depend on  changes in the incident wavenumber, for linear and triangular clusters respectively. Figures\ref{sens1} - \ref{sens4} illustrate  the influence of changes in scatterer positions (Figures \ref{sens1a} - \ref{sens4a}) and impedances (real and imaginary parts, Figures \ref{sens1b} - \ref{sens4b}) on reflection and transmission coefficients for the linear and triangular cluster.}

\rev{These sensitivity studies show that  dependence on the  horizontal ($x$-)positions of the grating scatterers are negligible. They also indicate that linear clusters display small sensitivity to the positioning of the central scatterer.
} \an{ The first of linear clusters (\circled{1}), i.e. a linear configuration with small scatterers' spacing, shows sensitivity measures related to position changes that are two orders smaller than for other clusters (linear \circled{2} and triangular \circled{1} and \circled{2}). The target transmitted mode coefficient, $t_{-1}$, is the least sensitive parameter to changes in scatterers' positions (see figs. \ref{sens1a} - \ref{sens4a}), meaning that grating performance will be rather affected by increase in other diffracted mode amplitudes, than decrease in the target mode amplitude, $t_{-1}$.} 

\an{In general, it can be seen that for both cluster types, i.e. linear and triangular, small (negligible) variations in reflection/transmission coefficients are expected for small shifts of all scatterers' positions, as shown in figs. \ref{sens1a} to \ref{sens4a}. Also, provided impedance values given in tab. \ref{tabclusters}, relatively small impact of changes in impedances (real and imaginary parts) on reflection and transmission coefficients can be seen from figs. \ref{sens1b} - \ref{sens4b}. Interestingly, for both cluster types, damping properties are critical to the grating performance. For the two linear clusters considered (\circled{1} and \circled{2}) and the second configuration for the triangular cluster (\circled{2}), it is seen that damping - related to the imaginary part of impedance - has the highest influence on $t_{-1}$.}

\section{Summary}\label{summ}

We have described  a general approach for the inverse design of gratings for flexural waves in thin plates. Using a one-dimensional  periodic arrangement of clusters of a finite number of point attachments it is possible to channel the incident energy towards a desired direction. The general solution for the inverse problem requires  a cluster of both active and passive attachments, however it is possible to find solutions with only passive point scatterers.  The  required mechanical properties of the attached scatterers are defined by the impedances, which are obtained by solving a linear system of equations.  We have shown through specific examples that some configurations, the linear clusters, possess very  low dissipation, resulting in very high conversion to the desired refracted mode.  It should be noted that the impedances of the cluster elements are linearly related to the desired diffraction parameters; the design process requires only  a matrix inversion.  
It has to be pointed out that the present approach, although derived for flexural waves and for the specific example of the negative refractor, can be easily exported to other waves and devices.

\acknowledgments

ANN acknowledges support from the National Science Foundation under Award No. EFRI 1641078 and  the Office of Naval Research under MURI Grant No. N00014-13-1-0631.
 P.P. acknowledges support from the National Centre for Research and Development under the research programme LIDER (Project No. LIDER/317/L-6/ 14/NCBR/2015).
 D.T. acknowledges financial support through the ``Ram\'on y Cajal'' fellowship and by the U.S. Office of Naval Research under Grant No. N00014-17-1-2445.

\appendix   
\section{Plate Green's function} \label{A}  
The Green's function, which satisfies 
\beq{1=1}
D\big(\Delta ^2 G({\bf r}) - k^4 G({\bf r})\big)  = \delta ({\bf r}) ,
\eeq
can be readily obtained using a double Fourier transform as 
\beq{1=2}
G({\bf r}) = \frac 1{D(2\pi)^2} 
\int_{{\mathbb R}^2} \frac{e^{\ii(\xi x + \eta y)} \dd \xi \dd \eta  }
{ (\xi^2 + \eta^2)^2-k^4} .
\eeq
 Evaluating  the $\eta$ integral using the Cauchy residue theorem  gives
\bse{1=31}
\bal{1=3}
G({\bf r}) &= \frac 1{2\pi}
\int_{{\mathbb R}} \dd \xi\, e^{\ii \xi x  } f(\xi, y)   ,
\\
f(\xi, y) &= \frac 1{4Dk^2} 
\bigg( 
\frac{ e^{-(\xi^2-k^2)^{1/2} |y|} } {(\xi^2-k^2)^{1/2} }
- 
\frac{ e^{-(\xi^2+k^2)^{1/2} |y|} } {(\xi^2+k^2)^{1/2} }
\bigg) .\label{4=33}
\eal
\ese
Note that $(\xi^2-k^2)^{1/2} = -\ii \sqrt{k^2-\xi^2}$ for $|\xi|<k$. 
The explicit form \eqref{3} follows using known integral representations for the Hankel function.

The line sum
\beq{1=4}
\sum_{m\in \mathbb Z} e^{\ii k_x ma} \, 
G({\bf r}- ma \hat{\bf x} ) =  \sum_{m \in \mathbb Z} 
\frac 1{2\pi}
\int_{{\mathbb R}} \dd \xi\, e^{\ii  ma (k_x -\xi)  }  \, e^{\ii \xi x} f(\xi, y) 
\eeq
can be simplified using the Poisson summation formula
\beq{1=5}
 \sum_{m \in \mathbb Z}  \frac 1{2\pi}
\int_{{\mathbb R}} \dd u\, e^{\pm \ii  m u   } F(u)
=    \sum_{n \in \mathbb Z} F (2\pi n).  
\eeq
Hence, 
\beq{1=6}
\sum_{m\in \mathbb Z}  e^{\ii k_x ma} 
G({\bf r}- ma \hat{\bf x} ) =  \frac 1a \sum_{n \in \mathbb Z} 
e^{\ii (k_x + \frac{2\pi}a n)x} f(k_x + \frac{2\pi}a n, y)  
\eeq
with $f$ defined in \eqref{4=33}.  This gives the identity \eqref{151}.

\section{The $N=3$ retroreflector grating, linear cluster} \label{B} 

Assuming a configuration of $N=3$ scatterers, and using the fact that  
${\bf k}_{-1}^\pm = - {\bf k}_{0}^\mp$ for the negative refractor, 
\eqref{27.1} becomes 
\beq{27.2}
 \hat{\bf S} = \frac {\ii G_0}{k \sin \theta_0} 
\begin{pmatrix} 
    1
	& e^{-\ii {\bf k}_0^+ \cdot {\bf R}_2 } &  
	  e^{-\ii {\bf k}_0^+ \cdot {\bf R}_3 }            \\
	  1
	& e^{-\ii {\bf k}_0^- \cdot {\bf R}_2 } &  
	  e^{-\ii {\bf k}_0^- \cdot {\bf R}_3 }      \\ 
	 1
	& e^{\ii {\bf k}_0^+ \cdot {\bf R}_2 } &  
	  e^{\ii {\bf k}_0^+ \cdot {\bf R}_3 }    
\end{pmatrix}, 
\eeq
 Taking  ${\bf R}_2 = \bm R_+$, ${\bf R}_3 = \bm R_-$, where 
 $\bm R_\pm = \pm d(\cos\theta_d,\sin\theta_d)$, we have 
\beq{27.3}
 \hat{\bf S} = \frac {\ii G_0}{k \sin \theta_0} 
\begin{pmatrix} 
     1 & e^{-i \phi_{-}} & e^{ i \phi_{-}} \\
     1 & e^{-i \phi_{+}} & e^{ i \phi_{+}} \\
     1 & e^{ i \phi_{-}} & e^{-i \phi_{-}} \\
\end{pmatrix} 
\eeq
where $\phi_{\pm}=kd\cos(\theta_d\pm\theta_0)$.  Note that 
\bal{27.4}
\det \hat{\bf S} &= \frac {4G_0^3 }{(k \sin \theta_0 )^3 }
 \big( \cos\phi_- -\cos\phi_+ \big)  \sin \phi_-
\notag \\ 
&= -\Big(\frac {2G_0}{k \sin \theta_0} \Big)^3 
\sin \big( kd \cos(\theta_d - \theta_0) \big) \, 
\sin \big( kd \cos\theta_d \cos\theta_0 \big) \, 
\sin \big( kd \sin\theta_d \sin\theta_0 \big) 
\eal
which clearly vanishes at the "forbidden" values $\theta_d = 0$, $\frac {\pi}2$ and $\pi$.  However, referring to \eqref{735}, 
\beq{274}
\hat{\bf S}^{-1} {\bf e}_1 =  
\frac{ k \sin \theta_0    }
{ 4\ii G_0  \sin \phi_-   \sin \frac 12 (\phi_+-\phi_-)   }
\begin{pmatrix} 
     -2 \cos  \frac 12 (\phi_++\phi_-)  \\
		e^{ {\ii}(\phi_+ -\phi_-)/2 }\\
		e^{ {\ii} (\phi_- -\phi_+) /2}
\end{pmatrix}   
\eeq
which is well defined for $\theta_d = \frac {\pi}2$ even though $\det \hat{\bf S} =0$ at that angle.

%



\end{document}